\documentclass[12pt]{article}

\def\+{{+\!\!\!+}}
\def\pp{\mbox{\tiny${}_{\stackrel\+ =}$}}

\def\d{\partial}

\def\th{\theta}

\def\D{{\cal D}}

\def\P{\Phi}

\def\L{{\cal L}}

\def\p{\psi}

\def\Qp{{\bf Q}_+}
\def\Qm{{\bf Q}_-}
\def\Qpm{{\bf Q}_\pm}
\def\bR{\hbox{I\hspace{-0.04in}R}} 
\def\bid{\hbox{1\hspace{-0.04in}I}} 
\def\pmb#1{\setbox0=\hbox{#1}%
\kern.0em\copy0\kern-\wd0
\kern-.04em\copy0\kern-\wd0
\kern.08em\copy0\kern-\wd0
\kern-.04em\raise.0433em\box0 }         

\def\half{\frac{1}{2}}

\def\high{\vphantom{\half}}

\newcommand{\nc}{\newcommand}
\nc{\beq}{\begin{equation}}
\nc{\eeq}[1]{\label{#1}\end{equation}}
\nc{\ber}{\begin{eqnarray}}
\nc{\eer}[1]{\label{#1}\end{eqnarray}}
\nc{\pek}[1]{\cite{#1}}
\nc{\enr}[1]{(\ref{#1})}
\nc{\kal}[1]{{\cal{#1}}}
\nc{\dott}{\;\cdot\;}

\newcommand{\be}{\begin{equation}}
\newcommand{\ee}{\end{equation}}
\newcommand{\bea}{\begin{eqnarray}}
\newcommand{\eea}{\end{eqnarray}}

\newcommand{\Section}[1]{\section{#1} \setcounter{equation}{0}}

\begin{document}


\begin{center}
                       
                                \hfill   hep-th/0410217\\
                                \hfill   ITP-UH-24/04\\
                                \hfill   UUITP-24/04\\
                                \hfill   HIP-2004-54/TH\\

\vskip .3in \noindent

\vskip .1in

{\large \bf{T-duality for the sigma model with boundaries}}
\vskip .2in

{\bf Cecilia Albertsson}$^a$\footnote{e-mail
 address: cecilia@itp.uni-hannover.de},
{\bf Ulf Lindstr\"om}$^b$\footnote{e-mail address: ulf.lindstrom@teorfys.uu.se}
 and  {\bf Maxim Zabzine}$^{c}$\footnote{e-mail
 address: m.zabzine@qmul.ac.uk} \\

\vskip .15in

\vskip .15in
$^a${\em Institut f\"ur Theoretische Physik, Universit\"at Hannover, \\
Appelstra{\ss}e 2, DE-30167 Hannover, Germany}
\vskip .15in
$^b${\em  Department of Theoretical Physics, Uppsala University, \\
Box 803, SE-751 08 Uppsala, Sweden\\
and \\
HIP -- Helsinki Institute of Physics\\
P.O.\ Box 64 FIN-00014 University of Helsinki, Suomi -- Finland}\\
\vskip .15in
$^c${\em School of Mathematical Sciences, Queen Mary, University of London, \\
Mile End Road, London, E1 4NS, UK}

\bigskip

\vskip .1in

\end{center}
\vskip .4in

\begin{center} {\bf ABSTRACT } 
\end{center}
\begin{quotation}\noindent 
We derive the most general local boundary conditions necessary for
T-duality to be compatible with superconformal invariance of the
two-dimensional ${\cal N}$=1 supersymmetric nonlinear sigma model with
boundaries. To this end, we construct a consistent gauge invariant
parent action by gauging a $U(1)$ isometry, with and without boundary
interactions.  We investigate the behaviour of the boundary conditions
under T-duality, and interpret the results in terms of D-branes.
\end{quotation}
\vfill
\eject

\Section{Introduction}

Classical T-duality can be realised as a transformation acting on the
worldsheet fields in the two-dimensional nonlinear sigma model. This
realisation is straightforward when there is an isometry of the target
space which at the same time is a symmetry of the worldsheet action.
By gauging the isometry (making it locally dependent on the worldsheet
coordinates) one obtains a gauge invariant ``parent action'' from
which the original and dual actions are extracted by integrating out
different fields \cite{Lindstrom:1983rt,HullSpence,HullSpence2}. In
the process one finds the transformation on the worldsheet fields as
well as Buscher's rules \cite{Buscher1,Buscher2} which yield the dual
background.

The simplest situation occurs when the isometry group is abelian.  In
this case the duality transformation is symmetric in the sense that
one can start from either theory (the original or the dual one) and
obtain the other one via the same gauging process \cite{Alvarez2}.
But for nonabelian isometries the dual model does not possess the same
nonabelian isometry as the original theory, rendering the gauging
procedure less straightforward \cite{Ossa}. In such a situation, one
obtains the dual model by gauging a subgroup of the isometry, but a
corresponding gauging of the dual model does not bring you back to the
original model.

T-duality by way of gauging isometries has been extensively studied
for the nonlinear sigma model \cite{Ooguri,BorlafLozano,Hori}.  In
particular closed string T-duality in the presence of a $U(1)$
isometry is well understood, both for the bosonic model and the models
with ${\cal N}$=1 and ${\cal N}$=2 worldsheet supersymmetry
\cite{HKLR86,Rocek,Hassan2,Alvarez}. The operative point in those
investigations is that the parent action is gauge invariant \emph{up
to boundary terms}, if the background is subject to certain
constraints. For the open string, however, the worldsheet has
boundaries, and the gauging procedure must be performed with care so
as to make the parent action gauge invariant also on the boundary.  In
the bosonic case this is achieved by imposing the appropriate boundary
conditions on the worldsheet fields and on the added gauge fields. But
the supersymmetric action in general receives an extra boundary term
\cite{Haggi-Mani:2000uc,ALZ2}, necessitating an addition of gauge
fields on the boundary.  These additional boundary terms affect the
resulting boundary conditions.

The structure of the paper is as follows. In Section~\ref{scbc} we
review the ${\cal N}$=1 supersymmetric nonlinear sigma model with
boundaries and the set of boundary conditions required for
superconformal invariance.  Assuming there is a group acting on the
target space we derive in Section~\ref{isometry} the conditions
necessary for the model to be invariant under the action of that
group, first for the bosonic case and then for the supersymmetric one.
We find that the symmetry, when it exists, is global (i.e., the
infinitesimal parameter of the isometry transformation is independent
of the worldsheet coordinates), and that the group is an isometry of
the target space metric.  The derivation involves two invariance
checks: the action and the boundary conditions.  Armed with the full
set of boundary conditions for superconformal and isometry invariance
of the original model, we then construct the appropriate parent
action, first the bosonic one in Section~\ref{bosonic}, and then the
${\cal N}$=1 supersymmetric one in Section~\ref{susyTduality}.  For
each parent action we derive the conditions necessary for a consistent
retrieval of the original and dual models.  Gauge invariance is
checked for the action, the field equations, and for the boundary
conditions themselves.  We moreover verify that the supersymmetric
parent action is supersymmetric also on the boundary, given the most
general classical ansatz for the fermionic boundary conditions.  In
Section~\ref{couplings} we add boundary interactions to the sigma
model, showing how to incorporate them in the T-duality
transformation.  The model dual to a superconformal model is expected
to also be superconformal, hence the worldsheet fields of the dual
model must obey the corresponding superconformality conditions.  We
briefly discuss in Section~\ref{Tdualbc} how these ${\cal N}$=1
conditions transform under T-duality.  Finally,
Section~\ref{conclusions} contains our conclusions and a wishlist for
future research.

\Section{Superconformal boundary conditions}
\label{scbc}

Here we give a summary of the derivation in \cite{ALZ2} of boundary
conditions for the ${\cal N}$=1 supersymmetric nonlinear sigma
model.\footnote{The ${\cal N}$=2 supersymmetric nonlinear sigma model
was treated in \cite{LZ1,LZ2,Zabzine}.}
This model is given by the following action,
\beq
S= \int d^2\xi d^2\theta\,\, D_{+} \Phi^\mu D_{-} \Phi^\nu E_{\mu\nu}
(\Phi)
+\frac{i}{2}\int d\tau\,\, B_{\mu\nu} [\psi_{+}^\mu
\psi_{+}^\nu + \psi_-^\mu \psi_-^\nu ] \,,
\eeq{goodaction}
where $E_{\mu\nu}\equiv g_{\mu\nu} + B_{\mu\nu}$ denotes a background
of general (Riemannian) metric $g_{\mu\nu}$ and an antisymmetric
two-tensor $B_{\mu\nu}$ with field strength $H_{\mu\nu\rho}=
\half \d_{[\mu} B_{\nu\rho]}$.
The boundary term is required for ${\cal N}$=1 supersymmetry
to be preserved on the boundary.
In component form (\ref{goodaction}) reads
(for notation, see Appendix \ref{a:11susy})
\begin{eqnarray}
  \label{sigmamodel}
\nonumber S &=& \int\limits_{\Sigma} d^2\xi\,\,
  \left [ \d_\+ X^\mu \d_= X^\nu E_{\mu\nu}  +
    i \psi^\mu_+\nabla^{(+)}_- \psi_+^\nu
    g_{\mu\nu} + i \psi_-^\mu \nabla^{(-)}_+ \psi^\nu_- g_{\mu\nu} \right. \\
&& \left.
 + \frac{1}{2} \psi^\lambda_+ \psi_+^\sigma \psi_-^\rho \psi_-^\gamma
    {\cal R}^{-}_{\rho\gamma\lambda\sigma} \right ] \,,
\end{eqnarray}
where the worldsheet coordinates are
$\tau \in \bR$ and $\sigma \in [0, \pi]$, and the curvature ${\cal
  R}^{\pm}_{\rho\gamma\lambda\sigma}$ is defined as
\beq
{\cal R}^{\pm\mu}_{\,\,\,\sigma\rho\lambda} =
\Gamma^{\pm\mu}_{\,\,\,\lambda\sigma,\rho}
- \Gamma^{\pm\mu}_{\,\,\,\rho\sigma,\lambda}
+ \Gamma^{\pm\mu}_{\,\,\,\rho\gamma}
\Gamma^{\pm\gamma}_{\,\,\,\lambda\sigma}
- \Gamma^{\pm\mu}_{\,\,\,\lambda\gamma}
\Gamma^{\pm\gamma}_{\,\,\,\rho\sigma} \,,
\eeq{defcur}
with $\Gamma^{\pm\mu}_{\,\,\nu\rho}$ given by
\beq
\Gamma^{\pm\nu}_{\,\,\rho\sigma} \equiv \Gamma^{\nu}_{\,\,\rho\sigma} \pm
g^{\nu\mu} H_{\mu\rho\sigma}\,,
\eeq{definGammapm}
and $\nabla^{(\pm)}_{\pm}$ by
\beq
\nabla^{(+)}_{\pm}\psi_{+}^\nu = \partial_{\pp}\psi_{+}^\nu +
\Gamma^{+\nu}_{\,\,\rho\sigma}\d_{\pp} X^\rho
\psi_{+}^\sigma,\,\,\,\,\,\,\,\,\,\, \nabla^{(-)}_{\pm}\psi_{-}^\nu =
\partial_{\pp}\psi_{-}^\nu + \Gamma^{-\nu}_{\,\,\rho\sigma}\d_{\pp}
X^\rho \psi_{-}^\sigma \,.
\eeq{covdervferm}
In (\ref{sigmamodel}) we have used the equations of motion for the
auxiliary field $F^\mu_{+-}$,
\beq
F^\rho_{+-} + \Gamma^{-\rho}_{\,\,\lambda\sigma}
\psi_{+}^\lambda \psi_{-}^\sigma =0 \,.
\eeq{Feom}

To find the supersymmetric boundary conditions from
(\ref{goodaction}), we require the boundary terms
in the Euler-Lagrange equations to vanish simultaneously with those
in the supersymmetry variation.
The two boundary variations read (again using (\ref{Feom}))
\ber
\nonumber \delta S &=& i \int d\tau \left [ (\delta \psi_+^\mu \psi_+^\nu -
\delta \psi_-^\mu \psi_-^\nu)g_{\mu\nu} \right . \\
&& \left .+ \delta X^\mu (i\d_\+ X^\nu
E_{\nu\mu} - i\d_= X^\nu E_{\mu\nu} + \Gamma^-_{\nu\mu\rho} \psi_-^\nu
\psi_-^\rho - \Gamma^+_{\nu\mu\rho} \psi_+^\nu \psi_+^\rho) \right ] ,
\label{vargood} \\
\nonumber
\delta_s S &=& \epsilon^- \int d\tau \,\,\left [ \d_\+ X^\mu \psi_-^\nu
E_{\mu\nu} - \eta \psi_+^\mu \d_= X^\nu E_{\mu\nu} + \eta \d_\+ X^\mu
\psi_+^\nu B_{\mu\nu} + \d_= X^\mu \psi_-^\nu B_{\mu\nu} \right .
\\
 &&- \left . \frac{i}{3}\eta H_{\mu\nu\rho} \psi_+^\rho \psi_+^\mu
\psi_+^\nu - \frac{i}{3} H_{\mu\nu\rho} \psi_-^\rho \psi_-^\mu
\psi_-^\nu \right ] \,.
\eer{afterFeq}
Since the boundary relates left- and right-movers, we make
a general ansatz for the relation between $\psi_+^\mu$
and $\psi_-^\mu$. The goal is then to find the restrictions
on this ansatz arising from $\delta S= \delta_s S=0$.
The most general local fermionic boundary
condition\footnote{In this paper we only consider local boundary
conditions for the $(X,\psi)$ fields. Non-local boundary conditions for
bosonic sigma models and their properties under T-duality have been
considered in \cite{Vassilevich:2001at}.}
allowed (classically) by dimensional analysis is of the simple form
\begin{equation}
  \label{fermans}
  \psi^\mu_- = \eta R^\mu_{\,\,\,\nu} (X) \psi^\nu_+ \,,
\end{equation}
where $R^\mu_{\,\,\,\nu}(X)$ is a locally defined object which transforms
as a (1,1) tensor field under coordinate transformations, and
$\eta = \pm 1$ corresponds to the choice of spin structures.
The bosonic counterpart of (\ref{fermans}) is obtained 
by means of a supersymmetry transformation,\footnote{For an explicit
  component form of this transformation, see Eq.\ (\ref{compsusytr}).}
and reads
\begin{equation}
  \label{boscond}
  \partial_= X^\mu - R^\mu_{\,\,\,\nu}\partial_{+\!\!\!+} X^\nu
  + 2i(P^\sigma_{\,\,\,\rho}
  \nabla_\sigma R^\mu_{\,\,\,\nu} + P^\mu_{\,\,\,\gamma} g^{\gamma\delta}
  H_{\delta\sigma\rho} R^\sigma_{\,\,\,\nu})\psi_+^\rho \psi_+^\nu =0 \,,
\end{equation}
where $P^\mu_{\,\,\,\nu} \equiv (\delta^\mu_{\,\,\nu} +
R^\mu_{\,\,\,\nu})/2$.
The conditions (\ref{fermans}) and (\ref{boscond}) satisfy $\delta S= \delta_s
S=0$ if and only if $R^\mu_{\,\,\,\nu}$ obeys
\begin{equation}
  \label{RgRcond1}
  g_{\rho\sigma} = R^\mu_{\,\,\,\rho} g_{\mu\nu} R^\nu_{\,\,\,\sigma} \,,
\end{equation}
\beq
P^\rho_{\,\,\,\tau} R^\mu_{\,\,\,\sigma} g_{\mu\nu} \nabla_\rho
R^\nu_{\,\,\,\gamma} + P^\rho_{\,\,\,\sigma} R^\mu_{\,\,\,\gamma} g_{\mu\nu}
\nabla_\rho R^\nu_{\,\,\,\tau}+ P^\rho_{\,\,\,\gamma} R^\mu_{\,\,\,\tau}
g_{\mu\nu} \nabla_\rho R^\nu_{\,\,\,\sigma} + 4 P^\mu_{\,\,\,\tau}
P^\nu_{\,\,\,\sigma} P^\rho_{\,\,\,\gamma} H_{\mu\nu\rho} = 0 \,.
\eeq{ALZ320}

\subsection{Interpretation}
\label{interp}

These constraints imply some interesting geometrical properties of the
boundary. To make these properties precise, we define an operator that
projects vectors onto the Dirichlet directions, and we call it the Dirichlet
projector $Q^\mu_{\,\,\,\nu}$ (see, e.g.,~\cite{Fotopoulos}).
It satisfies $Q^2=Q$ and
\beq
R^\mu_{\,\,\,\rho} Q^\rho_{\,\,\,\nu}
=Q^\mu_{\,\,\,\rho} R^\rho_{\,\,\,\nu}=-Q^\mu_{\,\,\,\nu}\,.
\eeq{RQ}
The relations (\ref{RQ}) are motivated by the expectation that
the Dirichlet condition on the fermions reads (see, e.g.,~\cite{Polchinski96})
\beq
 \psi^i_- + \eta \psi^i_+ =0\,,
\eeq{psiDirichlet}
where $i$ labels the Dirichlet directions in some chosen coordinate system.
In covariant form it reads
\beq
 Q^\mu_{\,\,\,\nu} (\psi^\nu_- + \eta \psi^\nu_+) =0\,.
\eeq{Qpsi}
The projector complementary to $Q^\mu_{\,\,\,\nu}$
is the Neumann projector $\pi^\mu_{\,\,\,\nu}
\equiv \delta^\mu_{\,\,\nu}-Q^\mu_{\,\,\,\nu}$, which
projects vectors onto the Neumann directions and which satisfies
\beq
\pi^\mu_{\,\,\,\rho} P^\rho_{\,\,\,\nu}
= P^\mu_{\,\,\,\rho} \pi^\rho_{\,\,\,\nu} = P^\mu_{\,\,\,\nu}\,.
\eeq{piP}
Then, by contracting (\ref{RgRcond1})
and (\ref{ALZ320}) with $Q^\tau_{\,\,\,\lambda}$, we
find the conditions
\beq
\pi^\mu_{\,\,\,\rho} g_{\mu\nu} Q^\nu_{\,\,\,\sigma} =0
\eeq{pigQ}
and
\begin{equation}
  \label{piinteg1}
  \pi^\mu_{\,\,\,\gamma} \pi^\rho_{\,\,\,\nu}
  \pi^\delta_{\,\,\,[\mu , \rho]} = 0 \,.
\end{equation}
In a suitably chosen coordinate system (\ref{pigQ}) implies the
diagonalisation of the metric, in a sense decoupling Neumann from
Dirichlet directions.  Condition (\ref{piinteg1}) is the integrability
condition for $\pi^\mu_{\,\,\,\nu}$, implying that the D-brane must be
a maximal integral submanifold of the target space \cite{Yano1,Yano2}.

Next we contract
(\ref{boscond}) with $Q^\mu_{\,\,\,\nu}$, and use (\ref{piinteg1}),
to obtain
\beq
Q^\mu_{\,\,\,\nu}\d_\tau X^\nu = 0 \,.
\eeq{QdX}
This is just the Dirichlet condition for the end of the string,
confining it to the brane defined by
$\pi^\mu_{\,\,\,\nu}$. 
Furthermore, by contracting Eq.\ (\ref{ALZ320}) with $\pi^\mu_{\,\,\,\nu}$
one finds
\begin{equation}
  \label{piEpiR}
  \pi^\rho_{\,\,\,\mu} E_{\sigma\rho} \pi^\sigma_{\,\,\,\nu}
-  \pi^\rho_{\,\,\,\mu} E_{\rho\sigma} \pi^\sigma_{\,\,\,\lambda}
  R^\lambda_{\,\,\,\nu} =0\,.
\end{equation}
If (\ref{piEpiR}) holds, then condition (\ref{ALZ320})
is just the statement that
the field strength of the B-field on the brane coincides with the pullback
of the background field strength $H_{\mu\nu\rho}$ to the brane.
Note that for a spacefilling brane (when all directions are Neumann),
Eq.\ (\ref{piEpiR}) implies, schematically, $R = E^{-1}E^t$.

\subsection{1D superfield formalism}
\label{1Dformalism}

The boundary ansatz (\ref{fermans}) and its superpartner
(\ref{boscond}) are easier to handle in 1D superfield
formalism than in component form.
In the notation of Appendix~\ref{a:1D}, the two conditions
can be written as a single equation, as follows (see also \cite{Sevrin}),
\beq
 (D K^\mu +S^\mu) = R^\mu_{\,\,\,\nu} (K)  (D K^\nu -S^\nu) \,,
\eeq{DKSbc}
where the 1D superderivative $D$ satisfies $D^2=i\partial_\tau$.
Contracting (\ref{DKSbc}) with $Q^\mu_{\,\,\,\nu}$ we find
\beq
 Q^\mu_{\,\,\,\nu}(K)  DK^\nu = 0\,,
\eeq{QDK}
which encodes the two Dirichlet conditions (\ref{Qpsi}) and (\ref{QdX}).
It is equivalent to
\beq
 DK^\mu = \pi^\mu_{\,\,\,\nu} (K) DK^\nu \,.
\eeq{piDK}
We can try to deduce some information also about $S^\mu$,
by solving for it in (\ref{DKSbc}). The result is
\beq
 2 P^\mu_{\,\,\,\rho} \pi^\rho_{\,\,\,\nu} S^\nu = - \left(
\delta^\mu_{\,\,\nu}-R^\mu_{\,\,\,\nu} \right) DK^\nu \,,
\eeq{SinDK}
implying that, in the presence of a B-field,
$\pi^\mu_{\,\,\,\nu} S^\nu \neq 0$,
so $S^\mu$ has a nonzero Neumann part.
When the B-field vanishes, however, we see that $\pi^\mu_{\,\,\,\nu}
S^\nu = 0$, because then it follows from (\ref{piEpiR}) that
$(\delta^\mu_{\,\,\nu}-R^\mu_{\,\,\,\nu} )\pi^\nu_{\,\,\,\rho}=0$.

\subsection{Summary}

\begin{table}[ht]
\vspace{1cm}
\begin{tabular*}{\textwidth}{@{\extracolsep{\fill}}|ccc|}\hline
& $
\begin{array}{rcl@{\hspace{2cm}}l}
g_{\rho\sigma} - R^\mu_{\,\,\,\rho} g_{\mu\nu} R^\nu_{\,\,\,\sigma} &=&0 &
                                                     (\ref{RgRcond1}) \\
R^\mu_{\,\,\,\rho} Q^\rho_{\,\,\,\nu}
  =Q^\mu_{\,\,\,\rho} R^\rho_{\,\,\,\nu}&=& -Q^\mu_{\,\,\,\nu} & (\ref{RQ}) \\
\pi^\mu_{\,\,\,\rho} g_{\mu\nu} Q^\nu_{\,\,\,\sigma} &=&0 &  (\ref{pigQ}) \\
\pi^\mu_{\,\,\,\gamma} \pi^\rho_{\,\,\,\nu}
  \pi^\delta_{\,\,\,[\mu , \rho]}  &=&0 &              (\ref{piinteg1}) \\
  \pi^\rho_{\,\,\,\mu} E_{\sigma\rho} \pi^\sigma_{\,\,\,\nu}
  -\pi^\rho_{\,\,\,\mu} E_{\rho\sigma} \pi^\sigma_{\,\,\,\lambda}
  R^\lambda_{\,\,\,\nu} &=&0 &                           (\ref{piEpiR}) \\
 Q^\mu_{\,\,\,\nu}(K)  DK^\nu  &=&0 &                       (\ref{QDK})
\end{array}
$ & \\ \hline
\end{tabular*}
\caption{Boundary conditions for ${\cal N}$=1 superconformal invariance
of the nonlinear sigma model, listed in order of appearance.
We have assumed the ansatz (\ref{DKSbc}).}
\vspace{1cm}
\label{N1bc}
\end{table}

Given the ansatz (\ref{DKSbc}) we summarise in Table~\ref{N1bc} the
boundary conditions necessary for ${\cal N}$=1 superconformal
invariance of the nonlinear sigma model.  These conditions may
alternatively be obtained by considering the currents corresponding to
superconformal invariance on the boundary \cite{ALZ2,ALZ1}.

\emph{Remark:} We collect conditions in tables
such as this, at the end of the sections where we derive them. In most
cases any boundary condition referred to in the text can be found in
one of those tables.

\Section{Isometry}
\label{isometry}

Let $M$ be a manifold with metric $g$ and local coordinates $X^\mu$.
Let the group $G$ act on $M$, and let the group action be realised at the
infinitesimal level by vector fields $k^\mu_A(X)$
which generate the corresponding Lie algebra,
$$
\left[k_A, k_B \right] = f_{AB}^{\,\,\,\,\,\,\,\,\,C} k_C \,,
$$
where $f_{AB}^{\,\,\,\,\,\,\,\,\,C}$ are the structure constants.
Under the action of $G$ the coordinates transform as
\beq
\delta_k X^\mu = \epsilon^A k^\mu_A(X) \,,
\eeq{Xtransform}
where $\epsilon^A$ is the infinitesimal parameter of the
transformation, independent of $X^\mu$.
When $M$ is the target space of the nonlinear sigma model, we want to
know the conditions necessary for $G$ to be a symmetry of the model.

First note that $k^\mu_A$ is a Neumann vector,
\beq
 Q^\mu_{\,\,\,\nu} k^\nu_A =0\,,
\eeq{Qk0}
a fact we use frequently in this paper.
This can be seen from the Dirichlet condition in a suitably chosen
coordinate system where $X^n$ label Neumann directions and $X^i$ are
Dirichlet directions,
\beq
\d_\tau X^i =0 \, ,
\eeq{dtXi}
the solution of which describes a foliation,
\beq
X^i = c^i \, ,
\eeq{Xici}
for some constants $c^i$.  In the supersymmetric model, where the
Neumann projector $\pi^\mu_{\,\,\,\nu}$ is integrable, each constant
describes an integral submanifold.  There are now two different
questions to ask regarding invariance under the group $G$.  First,
when is the foliation as a whole invariant, i.e., when is
Eq.~(\ref{dtXi}) invariant?  Second, when is each individual
submanifold invariant, i.e., when is Eq.~(\ref{Xici}) invariant for
any chosen constant $c^i$?  The first question will be answered
in what follows.  To answer the second question, we apply the
variation $\delta_k X^i = \epsilon^A k^i_A$ to (\ref{Xici}) and
find
$$
k^i_A =0 \,,
$$
the covariant version of which is Eq.~(\ref{Qk0}).

We first analyse the bosonic model and then the supersymmetric one,
in each case checking invariance of the action and of any relevant
boundary conditions.  Invariance of the field equations
should follow from that of the action.  Nevertheless, as a check, we have
verified that this is indeed the case.

\subsection{Bosonic model}
\label{boseps}

\subsubsection{Action}

The variation of the bosonic action
\beq
S_{\mathrm{bos}}= \int d^2\xi \,\, \d_{\+} X^\mu \d_{=} X^\nu E_{\mu\nu}
\eeq{bosNLSM}
under (\ref{Xtransform}) is given by
\ber
\nonumber \delta_k S_{\mathrm{bos}}&=& \int d^2\xi \,\, \left[
\epsilon^A \, \d_{\+} X^\mu \d_{=} X^\nu \, \L_{k_A} E_{\mu\nu}
+ \d_{\+} \epsilon^A \, \d_{=} X^\mu  \, k^\nu_A E_{\nu\mu} \right. \\
&& \left. +\d_{\+} X^\mu \, \d_{=} \epsilon^A \, k^\nu_A E_{\mu\nu}
\right] \, .
\eer{bosisvar}
If the parameter $\epsilon^A$ is constant, i.e., $\d_{\pp} \epsilon^A=0$,
then (\ref{bosisvar}) vanishes up to a boundary term if the Lie
derivative (defined in Appendix~\ref{a:geom}) of the metric with respect to
$k^\mu_A$ satisfies
\beq
{\cal L}_{k_A} g_{\mu\nu} =0
\eeq{Lg0}
and that of the B-field satisfies
\beq
\L_{k_A} B_{\mu\nu} = \omega_{A\,[\nu,\mu]}
\eeq{domega}
for some one-form $\omega_A$.
Eq.~(\ref{Lg0}) is equivalent to invariance of the metric under
(\ref{Xtransform}), meaning that the transformation is an isometry
and $k^\mu_A$ are Killing vectors.  In the special case with a single
Killing vector lying along the $X^0$-direction ($k^\mu_A =
\delta^\mu_0$), this condition reads $g_{\mu\nu,0}=0$, i.e., the
metric is independent of the isometry direction.  In compliance with
common practice, we call this choice of coordinates \emph{adapted}.

With the background satisfying Eqs.~(\ref{Lg0}) and (\ref{domega}), the
variation (\ref{bosisvar}) reduces to the boundary term
\beq
\delta_k S_{\mathrm{bos}} = \int d\tau \, \left[  \, \epsilon^A\, 
\omega_{A\,\mu} \d_\tau X^\mu \right]_{\sigma=0,\pi} \,.
\eeq{bccomp}
Using the Dirichlet condition (\ref{QdX}),
this vanishes if and only if, for some scalar field $\alpha_A$,
\beq
\pi^\mu_{\,\,\, \nu} \omega_{A\,\mu} = \d_\nu \alpha_A
\eeq{omegacond}
on the boundary.
Note that this does not imply that $\L_{k_A} B_{\mu\nu}=0$, since
$\omega_A$ may be anything along the Dirichlet directions.
More specifically, plugging (\ref{omegacond}) into (\ref{domega}) yields
\beq
\L_{k_A} B_{\mu\nu}= Q^\lambda_{\,\,\, [\nu,\mu]} \omega_{A\,\lambda}
+ Q^\lambda_{\,\,\, [\nu} \d_{\mu]} \omega_{A\,\lambda} \,.
\eeq{LkE}

If $\epsilon^A$ is holomorphic ($\d_{=} \epsilon^A=0$ only), then the
variation (\ref{bosisvar}) vanishes up to a boundary term if the
Killing vectors satisfy
\beq
\nabla^{(+)}_{\mu} k^\nu_A =0 
\eeq{nablaplusk}
in the bulk.  The remaining boundary term reads
$$
\delta_k S_{\mathrm{bos}} =\int d\tau \, \left[ \,
\epsilon^A \, k^\mu_A E_{\mu\nu} \d_\tau X^\nu \right]_{\sigma=0,\pi} \,,
$$
which vanishes if and only if
\beq
k^\mu_A E_{\mu\nu} \pi^\nu_{\,\,\, \rho} = 0 
\eeq{kEpi}
on the boundary,
where we have again used (\ref{QdX}).
In general this may have a solution; however, if the action of $G$ is free 
and transitive, then there is no solution for this system unless
$\pi^\mu_{\,\,\,\nu} =0$.

If $\epsilon^A$ is antiholomorphic ($\d_{\+} \epsilon^A=0$ only) we
similarly find the bulk condition
\beq
\nabla^{(-)}_{\mu} k^\nu_A =0 
\eeq{nablaminusk}
and the boundary condition 
\beq
k^\mu_A E_{\nu\mu} \pi^\nu_{\,\,\, \rho} = 0 \,.
\eeq{piEk}
Again, there is no solution when $G$ acts freely and transitively.

Note that (\ref{nablaplusk}) and (\ref{nablaminusk}) are satisfied by
the chirality conditions on the Killing vectors in the WZW-model
\cite{ALZ3}.  Namely, consider a group manifold of a Lie group ${\cal
G}$, with isometry group ${\cal G}\times{\cal G}$ generated by left-
and right-invariant Killing vectors $l^\mu_A$ and $r^\mu_A$.  In the
case $\d_{\pp} \epsilon^A =0$, the sigma model is invariant under the
isometry transformation (\ref{Xtransform}) (up to boundary terms and
only if (\ref{Lg0}) and (\ref{domega}) hold), with $k^\mu_A$ a linear
combination of $l^\mu_A$ and $r^\mu_A$.  In fact, the model is
invariant with respect to both $l^\mu_A$ and $r^\mu_A$ separately, and
each vector gives rise to an (anti-)chiral current,
$$
{\cal J}^A_+ \equiv -r^A_\mu \d_\+ X^\mu\,,
\quad\quad
{\cal J}^A_- \equiv l^A_\mu \d_= X^\mu \,.
$$
The Killing vectors obey the Cartan-Maurer equations,
$$
\nabla^{(+)}_\rho r^\mu_A =0 \,,\quad\quad
\nabla^{(-)}_\rho l^\mu_A =0\,,
$$
which are precisely the equations (\ref{nablaplusk}) and (\ref{nablaminusk}).

Finally, if $\epsilon^A$ is arbitrary ($\d_{\pp} \epsilon^A \neq 0$)
there is no way to make the variation (\ref{bosisvar}) vanish in the
bulk, so this case is not allowed.

In conclusion, the bosonic action (\ref{bosNLSM}) is invariant under
the group $G$ if and only if the conditions in Table~\ref{bosisconds}
hold and the parameter $\epsilon^A$ is independent of the worldsheet
coordinates $\xi^\pm$, i.e., $\d_{\pp} \epsilon^A =0$; $G$ is then an
isometry group.  Eqs.~(\ref{QdX}) and~(\ref{omegacond}) are boundary
conditions, whereas Eqs.~(\ref{Qk0}), (\ref{Lg0}) and~(\ref{domega})
hold in the bulk.

\begin{table}[ht]
\vspace{1cm}
\begin{tabular*}{\textwidth}{@{\extracolsep{\fill}}|ccc|}\hline
& $
\begin{array}{rcl@{\hspace{2cm}}l}
Q^\mu_{\,\,\,\nu} \d_\tau X^\nu &=&0 &               (\ref{QdX}) \\
Q^\mu_{\,\,\,\nu} k^\nu_A &=&0 &                     (\ref{Qk0}) \\
\L_{k_A} g_{\mu\nu} &=&0 &                           (\ref{Lg0}) \\
\L_{k_A} B_{\mu\nu} - \omega_{A[\nu,\mu]} &=&0 &  (\ref{domega}) \\
\pi^\mu_{\,\,\,\nu} \omega_{A\,\mu}
   - \d_\nu \alpha_A &=& 0 &                   (\ref{omegacond})
\end{array} $ & \\ \hline
\end{tabular*}
\caption{Conditions for the bosonic action (\ref{bosNLSM})
to be invariant under the group $G$.}
\vspace{1cm}
\label{bosisconds}
\end{table}

\subsubsection{Boundary conditions}

Of the boundary conditions in Table~\ref{bosisconds} the only one
transforming nontrivially under (\ref{Xtransform}) is the Dirichlet
condition (\ref{QdX}).  Using Eq.~(\ref{Qk0}) its variation reads
$$
\epsilon^A k^\sigma_A \pi^\rho_{\,\,\,\sigma} \pi^\mu_{\,\,\,[\rho,\nu]}
\pi^\nu_{\,\,\,\lambda} \d_\tau X^\lambda =0 \,,
$$
implying that
\beq
k^\sigma_A \pi^\rho_{\,\,\,\sigma} \pi^\mu_{\,\,\,[\rho,\nu]}
\pi^\nu_{\,\,\,\lambda} =0 \,.
\eeq{kpiinteg}
Thus invariance of the boundary conditions requires the extra condition
(\ref{kpiinteg}).

We also need to check invariance of the boundary relation between
left- and right-moving worldsheet fields.
The most general classical boundary ansatz compatible with
conformal invariance of the bosonic model is
\beq
\d_= X^\mu = R^\mu_{\,\,\,\nu} \d_\+ X^\nu  \, ,
\eeq{bosans}
which is invariant under (\ref{Xtransform}) if
\beq
\L_{k_A} R^\mu_{\,\,\,\nu} =0 \, .
\eeq{LieR}
One can show that this implies
\beq
\L_{k_A} \pi^\mu_{\,\,\,\nu} = \L_{k_A} Q^\mu_{\,\,\,\nu} =0 \, .
\eeq{LieQpi}
Thus the conditions for the boundary ansatz (\ref{bosans})
to be invariant under $G$ are Eqs.~(\ref{LieR}) and (\ref{LieQpi}).

\subsubsection{Summary and interpretation}
\label{bosint1}

\begin{table}[ht]
\vspace{1cm}
\begin{tabular*}{\textwidth}{@{\extracolsep{\fill}}|ccc|}\hline
& $
\begin{array}{rcl@{\hspace{2cm}}l}
Q^\mu_{\,\,\,\nu} \d_\tau X^\nu &=&0 &                 (\ref{QdX}) \\
Q^\mu_{\,\,\,\nu} k^\nu_A &=&0 &                       (\ref{Qk0}) \\
\L_{k_A} g_{\mu\nu} &=&0 &                             (\ref{Lg0}) \\
{\cal L}_{k_A} B_{\mu\nu}- \d_{[\mu} \omega_{A \, \nu]}  &=& 0&
                                                    (\ref{domega}) \\
\pi^\nu_{\,\,\, \mu} \omega_{A\,\nu} - \d_\mu \alpha_A  &=& 0&
                                                 (\ref{omegacond}) \\
k^\sigma_A \pi^\rho_{\,\,\,\sigma} \pi^\mu_{\,\,\,[\rho,\nu]}
     \pi^\nu_{\,\,\,\lambda}               &=&0 & (\ref{kpiinteg}) \\
\d_= X^\mu - R^\mu_{\,\,\,\nu} \d_\+ X^\nu  &=&0&   (\ref{bosans}) \\
\L_{k_A} R^\mu_{\,\,\,\nu} &=& 0 &                    (\ref{LieR}) \\
\L_{k_A} \pi^\mu_{\,\,\,\nu}
     = \L_{k_A} Q^\mu_{\,\,\,\nu} &=& 0 &           (\ref{LieQpi})
\end{array} $ & \\ \hline
\end{tabular*} 
\caption{Conditions for invariance of the bosonic nonlinear sigma
model under the group $G$.}
\vspace{1cm}
\label{bosissum}
\end{table}

In Table~\ref{bosissum} we summarise the conditions necessary for the
bosonic model to be invariant under the group $G$.  These conditions
may be understood in the following way.  First, (\ref{QdX}) is just
the Dirichlet condition, defining the D-brane.  Second, (\ref{Qk0}) is
the statement that all the generating vectors $k^\mu_A$ are parallel
to the D-brane.  We have already seen that the bulk equation
(\ref{Lg0}) is the invariance of the metric under the transformation
(\ref{Xtransform}), implying that $G$ is an isometry group.  Together
with (\ref{LieQpi}) it implies
$$
\L_{k_A} \left( \pi^\mu_{\,\,\,\rho}
 g_{\mu\nu} \pi^\nu_{\,\,\,\sigma} \right) =0 \, .
$$
Moreover, Eq.~(\ref{LieR}) in adapted coordinates means that
$R^\mu_{\,\,\,\nu}$ is independent of the isometry direction $X^0$,
and similarly for (\ref{LieQpi}).  That is, the D-brane is an
invariant submanifold under the action of the group $G$ and it is
equipped with an invariant metric $\pi^\mu_{\,\,\, \rho} g_{\mu\nu}
\pi^\nu_{\,\,\,\sigma}$.

Next, Eq.~(\ref{domega}) is a restriction on the background in the
bulk which in adapted coordinates states that the dependence of the
B-field on the isometry direction is restricted (though not
necessarily zero).  The condition (\ref{omegacond}) says that the
components of $\omega_A$ parallel to the brane are restricted, but
we cannot say anything about the components orthogonal to it.
Eq.~(\ref{bosans}) is just the ansatz we made for the relation between
left- and right-movers on the worldsheet.  

Finally, we recognise (\ref{kpiinteg}) as the condition for
$\pi^\mu_{\,\,\,\nu}$ to be integrable, contracted with $k^\mu_A$,
cf.~Eq.~(\ref{piinteg1}).  Thus a restricted form of
$\pi$-integrability is a requirement for isometry invariance of the
bosonic action.  This yields a situation geometrically similar to that
in \cite{ALZ2}, implying that the projection of the brane onto the
isometry direction in the original model must be an integral
submanifold of the target space $M$.  Together with (\ref{domega}),
(\ref{omegacond}) and (\ref{LieQpi}), it implies that
$$
k^\lambda_A {\cal L}_{k_A} \left(
\pi^\mu_{\,\,\, \lambda} B_{\mu\nu} \pi^\nu_{\,\,\, \rho} \right) =0 \, ,
$$
i.e., the D-brane is equipped with a B-field
$\pi^\mu_{\,\,\, \lambda} B_{\mu\nu} \pi^\nu_{\,\,\,\rho}$ that is
$G$-invariant at least along the isometry directions.

\subsection{Supersymmetric model}
\label{susyeps}

\subsubsection{Action}

The supersymmetric
generalisation of the transformation (\ref{Xtransform}) reads
\beq
 \delta_k \Phi^\mu = \epsilon^A(\xi, \theta)  k^\mu_A(\Phi) \, ,
\eeq{transform}
where $k^\mu_A(\Phi)$ is a superfield whose lowest component is
$k^\mu_A(X)$, and $\epsilon^A(\xi, \theta)$ is a (superfield)
parameter that a priori depends on the worldsheet superspace
coordinates $(\xi^\pm, \theta^\pm)$.  The component form of
(\ref{transform}), which is needed to vary the boundary term in the
supersymmetric action (\ref{goodaction}), reads
\beq
\left\{ \begin{array}{lcl}
 \delta_k X^\mu &=& \epsilon^A(\xi) \, k^\mu_A(X) \, , \\
 \delta_k \psi^\mu_{\pm} &=& \epsilon^A_{\pm}(\xi) \,
k^\mu_A + \epsilon^A(\xi) k^\mu_{A\,\, ,\nu} \psi^\nu_{\pm}\, , \\
 \delta_k F^\mu_{+-} &=& \epsilon^A_{+-} (\xi) k^\mu_A
+ \epsilon^A_{+} (\xi) \, k^\mu_{A\,\, ,\nu}\psi^\nu_{-}
- \epsilon^A_{-} (\xi) \, k^\mu_{A\,\, ,\nu}\psi^\nu_{+} \\
&& + \epsilon^A(\xi)  \, k^\mu_{A\,\, ,\nu} F^\nu_{+-}
+ \epsilon^A(\xi) \, k^\mu_{A\,\, ,\nu\rho} \psi^\rho_{+}\psi^\nu_{-} \,,
\end{array} \right.
\eeq{comptrans}
where
$$
\epsilon^A (\xi) \equiv \epsilon^A(\xi, \theta)|_{\theta=0}\,, \quad\quad
\epsilon^A_{\pm}(\xi) \equiv (D_{\pm} \epsilon^A)\vert_{\theta=0}\,, \quad\quad
\epsilon^A_{+-} (\xi) \equiv (D_+D_- \epsilon^A)\vert_{\theta=0} \,.
$$
The variation of the action (\ref{goodaction})
under (\ref{transform}), (\ref{comptrans}) is given by
\ber
\nonumber \delta_k S&=& \int d^2\xi d^2\theta\,\, \left[
D_{+} \epsilon^A  \, D_{-} \Phi^\nu E_{\mu\nu} k^\mu_A 
 + D_{+} \Phi^\mu D_{-} \epsilon^A  \, E_{\mu\nu} k^\nu_A 
+ \epsilon^A  \, D_{+} \Phi^\mu D_{-} \Phi^\nu
 \L_{k_A} E_{\mu\nu} \right] \\
&& +\frac{i}{2}\int d\tau \,\,
\left( \epsilon^A \, \L_{k_A} B_{\mu\nu} \, [\psi_{+}^\mu
\psi_{+}^\nu + \psi_-^\mu \psi_-^\nu ]
 +2 k^\mu_A \, B_{\mu\nu} \,  [\epsilon^A_{+} \,
\psi_{+}^\nu  + \epsilon^A_{-} \, \psi_-^\nu ]
\right) \,.
\eer{isometryvar}
If the parameter $\epsilon^A$ is constant ($D_{\pm} \epsilon^A = 0$),
then $\delta_k S$ vanishes up to boundary terms if the background
satisfies (\ref{Lg0}) and (\ref{domega}).  The remaining boundary terms are
\begin{eqnarray*}
\delta_k S &=& i \int d^2\xi \,\,
\left[ - \partial_{\+} D_{-} \left( \epsilon^A \, \omega_{A\,\mu}
D_{-} \Phi^\mu \right)
+ \partial_{=} D_{+} \left( \epsilon^A \, \omega_{A\,\mu}
D_{+} \Phi^\mu \right) \right] \\
&&  +i \int d\tau\,\,
\epsilon^A \, \d_\mu \omega_{A\,\nu} \, [\psi_{+}^\mu
\psi_{+}^\nu + \psi_-^\mu \psi_-^\nu ]  \,.
\end{eqnarray*}
In components the fermion-terms cancel, and $\delta_k S$ reduces to
$$
\delta_k S = \int d\tau \, \left[  \, \epsilon^A \,
\omega_{A\,\mu} \d_\tau X^\mu \right]_{\sigma=0,\pi} \,.
\quad\quad \quad\quad (\ref{bccomp})
$$
Using the Dirichlet condition (\ref{QdX}),
this vanishes if and only if (\ref{omegacond}) holds on the boundary.
Consequently Eq.~(\ref{LkE}) holds, and 
using $\pi$-integrability (which is provided as a condition for
superconformal invariance) it follows that
\beq
\pi^\mu_{\,\,\, \rho} ( \L_{k_A}  B_{\mu\nu} ) \pi^\nu_{\,\,\, \sigma} =0\,.
\eeq{LkpiE}

If $\epsilon^A$ is (anti-) holomorphic, i.e., $D_+ \epsilon^A=0$ or
$D_-\epsilon^A=0$, then we find the same conditions as in the
corresponding bosonic case, see Eqs.~(\ref{nablaplusk})--(\ref{piEk}),
to which there are no solutions when $G$ acts freely and transitively.
If $\epsilon^A$ is arbitrary the variation (\ref{isometryvar}) does
not vanish.

In conclusion, the supersymmetric action (\ref{goodaction}) is
invariant under the group $G$ if and only if in addition to the
superconformality conditions in Table~\ref{N1bc} the conditions listed in
Table~\ref{actionis} are satisfied, and if the parameter $\epsilon^A$
is independent of the worldsheet supercoordinates
$(\xi^\pm,\theta^\pm)$, i.e., $D_{\pm} \epsilon^A = 0$. The group $G$
is then an isometry group, and the action is simultaneously
superconformal and isometry invariant.

\begin{table}[ht]
\vspace{1cm}
\begin{tabular*}{\textwidth}{@{\extracolsep{\fill}}|ccc|}\hline
& $
\begin{array}{rcl@{\hspace{2cm}}l}
Q^\mu_{\,\,\,\nu} \d_\tau X^\nu &=&0 &               (\ref{QdX}) \\
Q^\mu_{\,\,\,\nu} k^\nu_A &=&0 &                     (\ref{Qk0}) \\
\L_{k_A} g_{\mu\nu} &=&0 &                           (\ref{Lg0}) \\
\L_{k_A}  B_{\mu\nu} -\omega_{A[\nu,\mu]} &=&0 &  (\ref{domega}) \\
\pi^\mu_{\,\,\, \nu} \omega_{A\,\mu}
   - \d_\nu \alpha_A &=&0  &                   (\ref{omegacond}) \\
\pi^\mu_{\,\,\, \rho} ( \L_{k_A}  B_{\mu\nu} )
   \pi^\nu_{\,\,\, \sigma}&=&0 &                   (\ref{LkpiE})
\end{array} $ & \\ \hline
\end{tabular*}
\caption{Conditions for the supersymmetric action (\ref{goodaction}) to be
invariant under the group $G$.  We have assumed that the conditions listed in
Table~\ref{N1bc} for ${\cal N}$=1 superconformal invariance are satisfied.}
\vspace{1cm}
\label{actionis}
\end{table}

\subsubsection{Boundary conditions}

To verify invariance of the boundary conditions in the supersymmetric
model we work in 1D superfield formalism (see Appendix~\ref{a:1D}),
since this is much more convenient than working in components.  When
$D_\pm \epsilon^A =0$ the transformations (\ref{comptrans}) read
$$
\left\{ \begin{array}{lcl}
\delta_k X^\mu &=& \epsilon^A \,  k^\mu_A(X) \,, \\
\delta_k \psi^\mu_{\pm} &=& \epsilon^A \, k^\mu_{A,\nu} \psi^\nu_{\pm}\,,\\
\delta_k F^\mu_{+-} &=& \epsilon^A \,  k^\mu_{A,\nu} F^\nu_{+-} 
+ \epsilon^A \, k^\mu_{A\,\, ,\nu\rho} \psi^\rho_{+}\psi^\nu_{-} \,,
\end{array} \right.
$$
which in terms of 1D superfields reads
\ber
\left\{
\begin{array}{lcl}
 \delta_k K^\mu &=& \epsilon^A \,  k^\mu_A(K) \,, \\
 \delta_k S^\mu &=& \epsilon^A \, k^\mu_{A,\nu} (K) S^\nu \,,
\end{array} \right.
\eer{1Dtrans}
with $D\epsilon^A =0$. 

Of the boundary conditions in Tables~\ref{N1bc} and~\ref{actionis} the
only one transforming nontrivially under $G$ is the
Dirichlet condition (\ref{QDK}).  Its variation under (\ref{1Dtrans}),
using Eq.~(\ref{Qk0}), reads
$$
k^\sigma_A \pi^\rho_{\,\,\,\sigma} Q^\mu_{\,\,\,[\nu,\rho]}
\pi^\nu_{\,\,\,\lambda} DK^\lambda  =0\,,
$$
which is satisfied by virtue of $\pi$-integrability, Eq.~(\ref{piinteg1}).

Next we check invariance of the boundary ansatz (\ref{DKSbc}).  It is
invariant under (\ref{1Dtrans}) if and only if (\ref{LieR}) holds, and
as a consequence we have also (\ref{LieQpi}).  Note that in the case
of a spacefilling D-brane this is consistent with the condition
(\ref{LkpiE}) on the background, since then $\L_{k_A} B_{\mu\nu}$=0
and from the superconformality condition (\ref{piEpiR}) we know that
$R^\mu_{\,\,\,\nu}$=$E^{\mu\rho}E_{\nu\rho}$.

\subsubsection{Summary and interpretation}
\label{isometrysum1}

\begin{table}[ht]
\vspace{1cm}
\begin{tabular*}{\textwidth}{@{\extracolsep{\fill}}|ccc|}\hline
& $
\begin{array}{rcl@{\hspace{2cm}}l}
Q^\mu_{\,\,\,\nu} k^\nu_A &=&0 &                       (\ref{Qk0}) \\
\L_{k_A} g_{\mu\nu} &=&0 &                             (\ref{Lg0}) \\
{\cal L}_{k_A} B_{\mu\nu}- \d_{[\mu} \omega_{A \, \nu]}  &=& 0&
                                                    (\ref{domega}) \\
\pi^\nu_{\,\,\, \mu} \omega_{A\,\nu} - \d_\mu \alpha_A  &=& 0&
                                                 (\ref{omegacond}) \\
\L_{k_A} R^\mu_{\,\,\,\nu} &=& 0 &                    (\ref{LieR}) \\
\L_{k_A} \pi^\mu_{\,\,\,\nu}
     = \L_{k_A} Q^\mu_{\,\,\,\nu} &=& 0 &           (\ref{LieQpi}) \\ 
\pi^\mu_{\,\,\, \rho} ( \L_{k_A}  B_{\mu\nu} )
   \pi^\nu_{\,\,\, \sigma} &=& 0 &                   (\ref{LkpiE})
\end{array} $ & \\ \hline
\end{tabular*} 
\caption{Conditions for invariance of the ${\cal
N}$=1 supersymmetric sigma model under the group $G$.  We have assumed
that the superconformality conditions listed in Table~\ref{N1bc} are
satisfied.}
\vspace{1cm}
\label{issummary}
\end{table}

In Table~\ref{issummary} we summarise the conditions necessary for the
supersymmetric sigma model to be invariant under the group $G$, given the
superconformality conditions in Table~\ref{N1bc}.  These conditions
were all interpreted in Section~\ref{bosint1}, except Eq.~(\ref{LkpiE}).
This condition combined with (\ref{LieQpi}) implies
$$
\L_{k_A} (\pi^\mu_{\,\,\, \rho} B_{\mu\nu} \pi^\nu_{\,\,\, \sigma}) =0 \,.
$$
Together with the conclusions in Section~\ref{bosint1} and the
consequences of the conditions in Table~\ref{N1bc}, this means the
D-brane is a $G$-invariant maximal integral submanifold and it is
equipped with invariant geometrical data, namely an invariant metric and an
invariant two-form ($\pi^\mu_{\,\,\, \rho} g_{\mu\nu} \pi^\nu_{\,\,\,
\sigma}$ and $\pi^\mu_{\,\,\, \rho} B_{\mu\nu} \pi^\nu_{\,\,\,
\sigma}$, respectively).

\Section{Bosonic T-duality}
\label{bosonic}

Before studying the supersymmetric sigma model, we first highlight
some fundamental issues concerning gauging and T-duality that arise
already in the bosonic model. Most aspects of T-duality are well
understood in the absence of a boundary
\cite{HullSpence,HullSpence2,HKLR86,Rocek}. Here we use the gauging
procedure \cite{Alvarez2} to obtain the dual model via a parent action
with boundaries, determining the boundary conditions necessary for
this procedure to work. This analysis involves several checks. First,
the act of integrating out fields to obtain the original and dual
sigma models must be consistent, meaning that the parent action must
be gauge invariant through every step before the gauge is fixed.
Second, the boundary equations of motion must be gauge invariant as
well as mutually compatible. Third, the boundary conditions themselves
must be gauge invariant \cite{LRvN}. After completing these checks, we
next find the most general ansatz for the relation between left- and
right-movers and derive the conditions for this ansatz to be gauge
invariant.

We consider the simplest possible case, i.e., we gauge a $U(1)$ isometry
generated by a single Killing vector $k^\mu$.

\subsection{Consistent T-duality procedure}
\label{bosbulk}

Gauging the $U(1)$ isometry in the bosonic action (\ref{bosNLSM})
produces the following parent action,
\ber
\nonumber  S_P &=& \int d^2\xi \,\,\,
\left [(\d_{\+} X^\mu+k^\mu A_{\+})
 ( \d_= X^\nu + k^\nu A_{=})E_{\mu\nu} \right . \\
&& \left. +A_{=} \d_{\+} y - A_{\+} \d_= y 
+A_{=} \omega_\mu \d_\+ X^\mu
- A_{\+} \omega_\mu \d_= X^\mu \right ] \,.
\eer{bosgact0}
Here we have introduced independent gauge fields $A_{\pp}$ and a
Lagrange multiplier $y$. If the background satisfies (\ref{Lg0}) and
(\ref{domega}), then\footnote{Choosing a
gauge for $\omega$ such that ${\cal L}_k B_{\mu\nu}=\omega_{[\nu,\mu]}
=0$ is possible only when there is just one Killing vector $k^\mu$.
If the isometry group were, say, $U(1)\times U(1)$, with two commuting
Killing vectors $k^\mu_1, k^\mu_2$, then it would not be possible to
gauge away both Lie derivatives ${\cal L}_{k_{1,2}} B_{\mu\nu}$
simultaneously; see e.g.~\cite{HKLR86,DeWit}. This is true
for models without boundary.}
$S_P$ is invariant up to boundary terms under the following
gauge transformations,
\beq
\left\{ \begin{array}{lcl}
 \delta_k X^\mu &=& \epsilon(\xi) k^\mu(X) \,, \\
 \delta_k A_{\pp} &=& - \d_{\pp} \epsilon(\xi)\,, \\
 \delta_k y &=& -\epsilon(\xi) k^\mu(X) \omega_\mu(X) \,,
\end{array} \right . 
\eeq{bosgtransf}
where the gauge parameter $\epsilon$ is an unrestricted function
of the worldsheet coordinates. We can rewrite $S_P$ in terms of
gauge covariant quantities, i.e., we make the
substitution\footnote{These $\nabla_{\pp}$ operators
have nothing to do with the
$\nabla^{(\pm)}_{\pm}$ operators defined in (\ref{covdervferm}).}
$$
\nabla_{\pp} X^\mu \equiv \d_{\pp} X^\mu + k^\mu A_{\pp}\,,
$$
which transforms covariantly under (\ref{bosgtransf}),
$$
\delta_k \nabla_{\pp} X^\mu = \epsilon k^\mu_{\,\,,\nu} \nabla_{\pp} X^\nu \,.
$$
Then the parent action reads
\ber
\nonumber  S_P &=& \int d^2\xi \,\,\,
\left [\nabla_{\+} X^\mu \nabla_= X^\nu E_{\mu\nu} \right . \\
&& \left. +A_{=} \d_{\+} y - A_{\+} \d_= y 
+A_{=} \omega_\mu \nabla_\+ X^\mu - A_{\+} \omega_\mu \nabla_= X^\mu \right ] \,.
\eer{bosgact1}
This substitution will be useful in Section~\ref{bosansatz}
when discussing a boundary ansatz for the worldsheet fields.

\subsubsection{Equations of motion}
\label{boseom}

 From the parent action the original and dual sigma models are
obtained by integrating out the fields $y$ and $A_{\pp}$,
respectively, and then fixing the gauge. To integrate out the fields,
we need their equations of motion, both in the bulk and on the
boundary. The bulk equation of motion resulting from a variation of
$y$ is
\beq
\d_\+ A_{=} - \d_= A_{\+} =0\,,
\eeq{bosyeom}
implying that
\beq
A_{\pp} = \d_{\pp} \lambda
\eeq{bosyeom2}
for some scalar field $\lambda$. From varying $A_{\pp}$ we find
the bulk equations
\beq
\left\{
\begin{array}{l}
A_{\+} k^\mu k^\nu E_{\mu\nu}
+ \d_\+ X^\nu \left( k^\mu E_{\nu\mu} + \omega_\nu \right) + \d_\+ y = 0\,, \\
A_{=} k^\mu k^\nu E_{\mu\nu}
+ \d_= X^\nu \left( k^\mu E_{\mu\nu} - \omega_\nu \right) - \d_= y = 0 \,.
\end{array} \right.
\eeq{Aeom}
In addition, the variation with respect to $y$ produces a boundary term
(there are no boundary terms from varying $A_{\pp}$),
\beq
\int d\tau  \,\, \delta y \, (A_{\+} + A_{=}) =0\,.
\eeq{bosybeom}
To find the most general solution to (\ref{bosybeom})
that is compatible with the bulk
equation of motion (\ref{bosyeom2}), we restrict the latter
to the boundary and insert it in (\ref{bosybeom}).
The resulting condition reads
\beq
2\int d\tau  \,\, \delta y \, \d_\tau \lambda =0\,,
\eeq{bosybeom2}
which has two solutions: $\d_\tau y=0$ and $\d_\tau
\lambda=0$. The latter solution, however, is not allowed, since it
would entail a restriction of the gauge freedom of the action.  After
integrating out the fields, but before fixing the gauge, the action
must still be gauge invariant, otherwise we will not be able to use
the gauge transformations (\ref{bosgtransf}) to choose a gauge and
perform T-duality. Thus we are left with
\beq
\d_\tau y=0
\eeq{dty0}
as the most general solution to (\ref{bosybeom}).

\subsubsection{Recovering the original model}
\label{bosrec}

We may now proceed to integrate out $y$ by substituting
(\ref{bosyeom2}) and (\ref{dty0}) in the parent action
(\ref{bosgact1}). Having thus eliminated $y$ both in the bulk and on
the boundary, the resulting action reads
\ber
\nonumber  S_\lambda &=& \int d^2\xi \,\,\,
\left [(\d_{\+} X^\mu+k^\mu \d_{\+}\lambda)
 ( \d_= X^\nu + k^\nu \d_= \lambda)E_{\mu\nu} \right . \\
&& \left. +\d_= \lambda \, \omega_\mu \d_\+ X^\mu
- \d_{\+}\lambda \, \omega_\mu \d_= X^\mu \right ] \,.
\eer{bosnoy}
Before fixing the gauge we need to check that $S_\lambda$
is gauge invariant. It turns out that this requirement yields another
crucial boundary condition, which must be satisfied if the T-duality
operation is to be well-defined.  Given (\ref{Lg0}) and (\ref{domega}),
the transformation of $S_\lambda$ under
(\ref{bosgtransf}) yields the following boundary term,
\beq
\delta_k S_\lambda = 2\int d\tau \,\,\,
\epsilon \, \omega_\mu \left( \d_\tau X^\mu +k^\mu \d_\tau\lambda \right) \,.
\eeq{bosgnoy}
There are two possibilities for making this term vanish.
One is to require that the gauge parameter $\epsilon$
be zero on the boundary,
\beq
\epsilon \vert_{\sigma=0,\pi} =0\, .
\eeq{bose0}
However, this would restrict the gauge freedom of our action,
which, as remarked above, is not allowed. Thus the
only permissible condition that sets (\ref{bosgnoy}) to zero is
\beq
\omega_\mu  \left( \d_\tau X^\mu +k^\mu \d_\tau\lambda \right) =0\, .
\eeq{bosnoy1}
Using the Dirichlet condition $Q^\mu_{\,\,\,\nu} \d_\tau X^\nu=0$,
and since $\pi^\mu_{\,\,\,\nu} \d_\tau X^\nu$
is in general nonzero, the condition (\ref{bosnoy1}) is equivalent to
\beq
\pi^\mu_{\,\,\,\nu} \omega_\mu=0\, .
\eeq{piomega0}
That is, the Neumann components of $\omega_\mu$ vanish on the
boundary. Note that, since $Q^\mu_{\,\,\,\nu} k^\nu=0$, Eq.~(\ref{Qk0}),
this implies
$k^\mu \omega_\mu=0$, i.e., $\omega_\mu$ is orthogonal to the Killing
vector.

We conclude that the action (\ref{bosnoy}) is gauge invariant given
the two boundary conditions (\ref{dty0}) and
(\ref{piomega0}), and we can now safely proceed with the gauge-fixing.  We go
to adapted coordinates ($k^\mu = \delta^\mu_0$) and choose a
gauge where $\lambda=0$. The action $S_\lambda$ then reduces to the original
nonlinear sigma model,
\ber
S_{\mathrm{bos}} &=& \int d^2\xi \,\,\, \d_\+ X^\mu \d_= X^\nu E_{\mu\nu} \,.
\eer{bosorig}

\subsubsection{Finding the dual model}

To find the dual model we use (\ref{Aeom}) to integrate out $A_{\pp}$
in (\ref{bosgact1}). The resulting action reads
\ber
\nonumber  S_y &=& \int d^2\xi \,\,\,
\left [\d_{\+} X^\mu \d_= X^\nu E_{\mu\nu}
 + (k^\lambda k^\sigma E_{\lambda\sigma})^{-1} \times \right . \\
&& \left. \times \left(\d_\+ y +
\omega_\mu \d_\+ X^\mu + k^\mu E_{\nu\mu} \d_\+ X^\nu \right)
\left(\d_= y + \omega_\rho \d_= X^\rho
- k^\gamma E_{\gamma\rho} \d_= X^\rho \right)
 \right ] \,,
\eer{bosnoA}
which is required to be gauge invariant. Using (\ref{Lg0})
and (\ref{domega})
a gauge transformation leaves this action invariant up to the
following boundary term,
\beq
\delta_k S_y = 2\int d\tau \,\,\,
\epsilon \left( \d_\tau y + \omega_\mu \d_\tau X^\mu \right)\,.
\eeq{bosgnoA}
But we already know from Sections~\ref{boseom} and~\ref{bosrec}
that (\ref{dty0}) and
(\ref{piomega0}) must always hold. Hence (\ref{bosgnoA})
vanishes identically, and $S_y$ is indeed
gauge invariant. We can therefore
go to adapted coordinates
and choose the gauge where $X^0=0$. This yields the action
\beq
\widetilde{S}_{\mathrm{bos}}= \int d^2\xi \,\,
\d_\+ \widetilde X^\mu \d_= \widetilde X^\nu \widetilde{E}_{\mu\nu} \,,
\eeq{bosdual}
where
$$
\widetilde X^0 \equiv y \, , \quad\quad
\widetilde X^n \equiv X^n \, , 
$$
and $\widetilde{E}_{\mu\nu}$ is the dual background given by the
following (modified) Buscher's rules \cite{Buscher1,Buscher2,Alvarez2},
\beq
\left\{ \begin{array}{lcl}
\widetilde{E}_{00} &=& E^{-1}_{00}\,,  \\
\widetilde{E}_{0n} &=& -(E_{0n}-\omega_n) E^{-1}_{00}\,,  \\
\widetilde{E}_{m0} &=& (E_{m0}+\omega_m) E^{-1}_{00}\,,  \\
\widetilde{E}_{mn} &=&
   E_{mn}-(E_{m0}+\omega_m)(E_{0n}-\omega_n) E^{-1}_{00}\,.  \\
\end{array} \right.
\eeq{Buscher}
Thus we have found the dual action, Eq.~(\ref{bosdual}), which has
the same form as the original one, with $y$ playing
the part of $X^0$-coordinate.

\subsubsection{Summary}

In Table~\ref{Tconsist} we collect the boundary conditions necessary
for the bosonic parent action (\ref{bosgact1}) to give rise to the original
and dual nonlinear sigma models in a consistent manner.

\begin{table}[ht]
\vspace{1cm}
\begin{tabular*}{\textwidth}{@{\extracolsep{\fill}}|ccc|}\hline
& $
\begin{array}{rcl@{\hspace{2cm}}l}
Q^\mu_{\,\,\,\nu} \d_\tau X^\nu&=&0 &                      (\ref{QdX}) \\
Q^\mu_{\,\,\,\nu} k^\nu&=&0 &                              (\ref{Qk0}) \\
{\cal L}_{k} g_{\mu\nu} &=&0 &                             (\ref{Lg0}) \\
{\cal L}_{k} B_{\mu\nu} -\d_{[\mu} \omega_{\nu]} &=&0 & (\ref{domega}) \\
\d_\tau y &=&0 &                                          (\ref{dty0}) \\
\pi^\mu_{\,\,\,\nu} \omega_\mu &=&0 &                 (\ref{piomega0})
\end{array}
$ & \\ \hline
\end{tabular*}
\caption{Conditions for the bosonic parent action (\ref{bosgact1})
to consistently yield the original and dual nonlinear sigma models.}
\vspace{1cm}
\label{Tconsist}
\end{table}

\subsection{Gauge invariance of parent action}
\label{bosginv}

We know that the parent action is gauge invariant on the boundary as
well as in the bulk after integrating out the lagrange multiplier $y$
and gauge fields $A_{\pp}$, respectively, given the conditions in
Table~\ref{Tconsist}. Let us nevertheless make sure that the parent
action is gauge invariant also before the fields are integrated
out. This is quickly done: a gauge transformation of the parent action
(\ref{bosgact1}) leaves precisely the boundary term (\ref{bosgnoA}),
which vanishes identically due to (\ref{QdX}), (\ref{dty0}) and
(\ref{piomega0}).

\subsection{Compatibility of field equations}
\label{bosbdy}

Here we check that the equations of motion derived from the parent
action are consistent on the boundary. This means checking that all
boundary field equations are mutually compatible as well as compatible
with the restriction of bulk equations to the boundary.  We already
have the equations of motion for $y$ and $A_{\pp}$, so let us derive the
one remaining boundary field equation, namely that of $X^\mu$. A field
variation with respect to $X^\mu$ produces the following equation,
\beq
 \int d\tau \,\, \delta X^\mu \,\, \left[ (\d_= X^\nu + k^\nu
A_{=})E_{\mu\nu} -(\d_\+ X^\nu + k^\nu A_{\+})E_{\nu\mu} + (A_{\+} +
A_{=})\omega_\mu \right] =0\,,
\eeq{bosXbeom}
implying
\beq
\pi^\mu_{\,\,\,\rho} \left[ (\d_= X^\nu + k^\nu A_{=})E_{\mu\nu} -(\d_\+
X^\nu + k^\nu A_{\+}) E_{\nu\mu} \right] + \pi^\mu_{\,\,\,\rho} \omega_\mu
(A_{\+} + A_{=}) =0\,.
\eeq{bosXbeom2}
This is compatible with restricting the bulk equations of motion
(\ref{bosyeom}) and (the difference between) Eqs.~(\ref{Aeom}) to
the boundary if and only if the following boundary condition holds,
\beq
 \d_\tau y + \omega_\mu
\pi^\mu_{\,\,\,\nu} \left( \d_\tau X^\nu + k^\nu \d_\tau \lambda
\right)=0\,.
\eeq{bosybc}
Inserting (\ref{dty0}) and (\ref{piomega0}), the left-hand side
vanishes identically, showing that the $X^\mu$ boundary equation of
motion is compatible with the $y$ boundary equation of motion and the
bulk equations of motion for $A_{\pp}$ and $y$. Hence the parent action
(\ref{bosgact1}) is consistent at the level of field equations, given
the boundary conditions in Table~\ref{Tconsist}.

\subsection{Gauge invariance of boundary conditions}
\label{s:bosginvbc}

For complete consistency, we must also check that all the boundary
conditions we have used in this section are themselves gauge invariant.
We list them here:
$$
\begin{array}{rcl@{\hspace{1.2cm}}l}
Q^\mu_{\,\,\,\nu}\d_\tau X^\nu &=&0 &                     (\ref{QdX}) \\
Q^\mu_{\,\,\,\nu} k^\nu&=&0 &                             (\ref{Qk0}) \\
{\cal L}_{k} g_{\mu\nu} &=&0 &                            (\ref{Lg0}) \\
{\cal L}_{k} B_{\mu\nu} -\d_{[\mu} \omega_{\nu]}&=&0 & (\ref{domega}) \\
\d_\+ A_{=} - \d_= A_{\+}  &=&0                          & (\ref{bosyeom}) \\
A_{\+} k^\mu k^\nu E_{\mu\nu}
+ \d_\+ X^\nu \left( k^\mu E_{\nu\mu} + \omega_\nu \right)
   + \d_\+ y  &=&0                           & (\ref{Aeom})\mathrm{a} \\
A_{=} k^\mu k^\nu E_{\mu\nu}
+ \d_= X^\nu \left( k^\mu E_{\mu\nu} - \omega_\nu \right)
   - \d_= y   &=&0                           & (\ref{Aeom})\mathrm{b} \\
\d_\tau y &=&0                                         & (\ref{dty0}) \\
\pi^\mu_{\,\,\,\nu} \omega_\mu &=&0                &   (\ref{piomega0}) \\
\pi^\mu_{\,\,\,\rho} \left[ (\d_= X^\nu + k^\nu A_{=})E_{\mu\nu} -(\d_\+
X^\nu + k^\nu A_{\+}) E_{\nu\mu} \right] + \pi^\mu_{\,\,\,\rho} \omega_\mu
(A_{\+} + A_{=})  &=&0  &                                (\ref{bosXbeom2})
\end{array}
$$
The requirements for gauge invariance of each of these conditions are
as follows:
\vspace{-10pt}
\begin{enumerate}
\item Eqs.~(\ref{Qk0}), (\ref{Lg0}), (\ref{domega}), (\ref{bosyeom})
      and (\ref{piomega0}) are trivially gauge invariant.
\item (\ref{QdX}) is gauge invariant if (\ref{kpiinteg}) holds.
\item Eqs.~(\ref{Aeom}) are gauge invariant due to (\ref{domega}).
\item (\ref{dty0}) is gauge invariant if $k^\mu \omega_\mu =0$, which
      follows from (\ref{piomega0}).
\item Given (\ref{piomega0}), (\ref{bosXbeom2}) is gauge invariant if
      $Q^\rho_{\,\,\,\mu} {\cal L}_k \pi^\mu_{\,\,\,\nu}=0$,
      which is equivalent to  Eq.~(\ref{kpiinteg}).
\end{enumerate}
In conclusion, our boundary conditions are gauge invariant
provided (\ref{kpiinteg}) holds.

\subsection{Boundary ansatz}
\label{bosansatz}

We now turn to the ansatz for the relation between left- and
right-movers on the boundary. This is needed for conformal invariance
of the gauged model, and provides us with
boundary conditions in terms of a gluing matrix.
In the bosonic ungauged model, the most general ansatz reads
$$
\d_= X^\mu = R^\mu_{\,\,\,\nu} \d_\+ X^\nu \, .
\hspace{3cm} (\ref{bosans})
$$
In the gauged model, however, it turns out that the appropriate
ansatz is
\beq
\nabla_= X^\mu = R^\mu_{\,\,\,\nu} \nabla_\+ X^\nu \, .
\eeq{bosgans}
This is supported by an analysis of the conformal current
(the stress-energy tensor) of the gauged model. For the model to be
conformal on the boundary, the left- and right-moving components
$T_{\pm\pm}$ must satisfy
$$
0=T_{++} -T_{--} \equiv
  \nabla_\+ X^\mu g_{\mu\nu} \nabla_\+ X^\nu
+ \nabla_= X^\mu g_{\mu\nu} \nabla_= X^\nu \, .
$$
The most general solution to this is obtained by inserting
(\ref{bosgans}), which leads to the condition (\ref{RgRcond1}), i.e.,
$R^\mu_{\,\,\,\nu}$ must preserve the metric, the same condition
required for the ungauged model to be conformally invariant.  If we
were to use (\ref{bosans}) instead, we would in addition to
(\ref{RgRcond1}) find additional, very restrictive conditions,
hence that solution would be less general.

What other boundary conditions, in addition to (\ref{RgRcond1}),
follow from the ansatz (\ref{bosgans})?  Gauge invariance of
(\ref{bosgans}) requires that $\L_{k} R^\mu_{\,\,\,\nu}=0$,
Eq.~(\ref{LieR}), which implies
$\L_{k}\pi^\mu_{\,\,\,\nu}$=$\L_{k}Q^\mu_{\,\,\,\nu}=0$,
Eq.~(\ref{LieQpi}).  Two further conditions are obtained by inserting
(\ref{bosgans}) into the equation of motion (\ref{bosXbeom2}), and
using $\pi^\mu_{\,\,\, \nu} \omega_{\mu}$=0, Eq.~(\ref{piomega0}). The
result is Eq.~(\ref{pigQ}) (diagonalisation of the metric) and
Eq.~(\ref{piEpiR}), which arose in the analysis of the ungauged
supersymmetric model in \cite{ALZ2,ALZ1}, but here they arise as
conformality conditions in the gauged \emph{bosonic} model.

One may ask if the gluing matrix $R^\mu_{\,\,\,\nu}$ of the gauged model
is the same object as the gluing matrix in the ungauged model,
something that is not a priori clear. We can see that in fact this
must be the case, by the following simple observation. The current
associated to conformal symmetry of the gauged model has a form
identical to that of the original model, with the original fields
replaced by gauge covariant ones. As a consequence the analysis of
conformal invariance of the parent action is identical to that of the
ungauged action.  Hence the resulting conditions for the gluing matrix
of the gauged model have the same form as those of the original
model. These conditions should reduce to those of the original model
when going to adapted coordinates and fixing the gauge. The question
is then whether there is a generalisation of the gluing matrix,
depending on more fields than $X^\mu$, which reduces to
$R^\mu_{\,\,\,\nu}$ once we integrate out $y$ and set $k^\mu=
\delta^\mu_0$ and $A_{\pp}=0$.  The answer is no: for conformal reasons,
the only field the gluing matrix can depend on other than $X^\mu$ is
$y$, but there is no conformally allowed modification of
$R^\mu_{\,\,\,\nu}$ that vanishes when we use the $y$ field equations
and fix the gauge. On the other hand, the ansatz (\ref{bosgans})
reduces in a trivial way to the ansatz (\ref{bosans}) of the original
model.

\subsection{Summary and interpretation}
\label{bosinterp}

\begin{table}[ht]
\vspace{1cm}
\begin{tabular*}{\textwidth}{@{\extracolsep{\fill}}|ccc|}\hline
& $
\begin{array}{rcl@{\hspace{2cm}}l}
g_{\rho\sigma} - R^\mu_{\,\,\,\rho} g_{\mu\nu} R^\nu_{\,\,\,\sigma} &=&0 &
                                                        (\ref{RgRcond1}) \\
\pi^\mu_{\,\,\,\rho} g_{\mu\nu} Q^\nu_{\,\,\,\sigma} &=&0 & (\ref{pigQ}) \\
Q^\mu_{\,\,\,\nu}\d_\tau X^\nu &=&0 &                        (\ref{QdX}) \\
\pi^\rho_{\,\,\,\mu} E_{\sigma\rho} \pi^\sigma_{\,\,\,\nu}
  -\pi^\rho_{\,\,\,\mu} E_{\rho\sigma} \pi^\sigma_{\,\,\,\lambda}
  R^\lambda_{\,\,\,\nu} &=&0 &                            (\ref{piEpiR}) \\
Q^\mu_{\,\,\,\nu} k^\nu&=&0 &                                (\ref{Qk0}) \\
{\cal L}_{k} g_{\mu\nu} &=&0 &                               (\ref{Lg0}) \\
{\cal L}_k B_{\mu\nu} - \omega_{[\nu,\mu]} &=&0  &        (\ref{domega}) \\
k^\mu \pi^\rho_{\,\,\,\mu} \pi^\nu_{\,\,\,\gamma}
     \pi^\lambda_{\,\,\,[\rho,\nu]}&=&0  &              (\ref{kpiinteg}) \\
\L_{k_A} R^\mu_{\,\,\,\nu}  &=&0 &                          (\ref{LieR}) \\
\L_{k_A} \pi^\mu_{\,\,\,\nu}
     = \L_{k_A} Q^\mu_{\,\,\,\nu} &=& 0 &                 (\ref{LieQpi}) \\
\d_\tau y &=&0  &                                           (\ref{dty0}) \\
\pi^\mu_{\,\,\,\nu} \omega_\mu &=&0  &                  (\ref{piomega0}) \\
\nabla_= X^\mu - R^\mu_{\,\,\,\nu} \nabla_\+ X^\nu &=&0& (\ref{bosgans})
\end{array} $ & \\ \hline
\end{tabular*} 
\caption{Conditions for the bosonic nonlinear sigma model
to be consistent under T-duality.}
\vspace{1cm}
\label{Tbossum}
\end{table}

In Table~\ref{Tbossum} we summarise the boundary conditions necessary
for the bosonic nonlinear sigma model to be consistent under
T-duality.  The first four of these conditions were found and
interpreted already in Section~\ref{scbc}: (\ref{RgRcond1}) is the
preservation of the metric, (\ref{pigQ}) is the diagonalisation of the
metric, (\ref{QdX}) is the Dirichlet condition in the original model,
and (\ref{piEpiR}) implies indirectly that
the B-field on the D-brane is the pullback
of the background B-field.

Conditions (\ref{Qk0}), (\ref{Lg0}), (\ref{domega}), (\ref{kpiinteg}),
(\ref{LieR}) and (\ref{LieQpi}) were interpreted already in
Section~\ref{bosint1}.  Together with (\ref{piomega0}) these
conditions tell us what we already knew about the D-brane in the
original model, namely that it is an isometry invariant submanifold
equipped with an invariant metric and two-form, whose projection onto
the isometry direction is an integral submanifold.

Next, Eq.~(\ref{dty0}) is a Dirichlet condition for $y$, implying that
the brane on the dual side is transverse to the $y$-direction. This is
consistent with the expectation that T-duality parallel to a D-brane
decreases its dimension by one. The condition (\ref{piomega0}) means
that $\omega_\mu$ has no components parallel to the brane in the
original sigma model, but we cannot say anything about the components
orthogonal to it. Eq.~(\ref{bosgans}) is just the ansatz we made for
the relation between left- and right-movers on the worldsheet.

\Section{Supersymmetric T-duality}
\label{susyTduality}

The bosonic discussion generalises in most respects straightforwardly
to the supersymmetric model. We follow the same logic, starting with
consistency of the T-duality procedure, then checking gauge invariance
and compatibility of field equations. Here, however, we find that a
boundary ansatz for the worldsheet fields is needed at an early stage
to finish the analysis of the field equations. We therefore make a
brief digression at the appropriate time to discuss the ansatz, and
then complete the consistency check. In addition, we check gauge
invariance of the boundary conditions and verify that the action as
well as all boundary conditions are compatible with supersymmetry.

\subsection{Consistency of T-duality procedure}
\label{susybulk}

Gauging the $U(1)$ isometry in the bulk of the supersymmetric action
(\ref{goodaction}) produces the following parent action,
\ber
\nonumber  S_P &=& \int d^2\xi d^2\theta\,\,\,
\left[ \nabla_+\Phi^\mu \nabla_- \Phi^\nu E_{\mu\nu}(\Phi)
+\omega_\mu  D_{(+} \Phi^\mu \, V_{-)} +  D_{(-} Y \, V_{+)} \right] \\
&& +\frac{i}{2}\int d\tau\,\,  B_{\mu\nu} [\psi_{+}^\mu
\psi_{+}^\nu + \psi_-^\mu \psi_-^\nu ]\, ,
\eer{gaugeact0}
where\footnote{These $\nabla_\pm$ operators are unrelated to
the $\nabla^{(\pm)}_{\pm}$ operators defined in (\ref{covdervferm}).}
$$
\nabla_\pm \Phi^\mu \equiv D_\pm \Phi^\mu+k^\mu V_\pm\,,
$$
and the independent superfields $V_\pm$ and $Y$ are, respectively,
gauge fields and a Lagrange multiplier.  Given (\ref{Lg0}) and
(\ref{domega}), $S_P$ is invariant up to
boundary terms under the following gauge transformations,
\beq
\left\{ \begin{array}{rcl}
 \delta_k \Phi^\mu &=& \epsilon(\xi, \theta)  k^\mu(\Phi) \,, \\
 \delta_k V_{\pm} &=& - D_{\pm} \epsilon (\xi, \theta) \,, \\
 \delta_k Y &=& -\epsilon (\xi, \theta) k^\mu(\Phi) \omega_\mu(\Phi) \,.
\end{array} \right . 
\eeq{trangasusy}
One may now ask if not the boundary term in (\ref{gaugeact0}) should
also be modified in some way, to make the full action gauge invariant
and consistent also on the boundary.  First we note that leaving the
boundary term unchanged leads to very restrictive conditions on the
background, e.g., $k^\mu B_{\mu\nu}\pi^\nu_{\,\,\,\rho}=0$, implying that
a modification is indeed necessary. However, it is not immediately
clear what this modification should look like. We make the natural
choice to replace the worldsheet fermions $\psi^\mu_\pm$ with their
gauge covariant counterparts, defined by
$$
\widehat\psi^\mu_\pm \equiv \nabla_\pm\Phi^\mu \vert_{\theta=0}
= \psi^\mu_\pm + k^\mu v_\pm\,,
$$
where
$$
v_\pm \equiv V_\pm |_{\theta=0} \,.
$$
With this modification of the boundary term, the gauged action now reads
\ber
\nonumber  \widehat S &=& \int d^2\xi d^2\theta\,\,\,
\left[ \nabla_+\Phi^\mu \nabla_- \Phi^\nu E_{\mu\nu}(\Phi)
+\omega_\mu  D_{(+} \Phi^\mu \,\, V_{-)} +  D_{(-} Y \,\, V_{+)} \right] \\
&& +\frac{i}{2}\int d\tau\,\,  B_{\mu\nu} [\widehat\psi_{+}^\mu
\widehat\psi_{+}^\nu + \widehat\psi_-^\mu \widehat\psi_-^\nu ] \,.
\eer{gaugeact1}
This is the action we will work with, and we will see
that it is consistent and supersymmetric. We begin by showing that it
gives rise to the appropriate original and dual sigma models in a
consistent way.

\subsubsection{Equations of motion}
\label{eom}

To obtain the original and dual models we need to integrate out $Y$
and $V_{\pm}$, respectively, by using their field equations in
(\ref{gaugeact1}), and then fix the gauge.  A variation with respect to
$Y$ yields the following bulk equation of motion,
\beq
D_+V_- + D_-V_+ =0,
\eeq{Yeom}
implying
\beq
V_\pm = D_\pm \Lambda
\eeq{Yeom2}
for some scalar superfield $\Lambda$.
Varying $V_{\pm}$ yields the field equations
\beq
\left\{
\begin{array}{l}
V_+ k^\mu k^\nu E_{\mu\nu}
+ D_+ \Phi^\nu \left( k^\mu E_{\nu\mu} + \omega_\nu \right) + D_+ Y = 0\,, \\
V_- k^\mu k^\nu E_{\mu\nu}
+ D_- \Phi^\nu \left( k^\mu E_{\mu\nu} - \omega_\nu \right) - D_- Y = 0\, .
\end{array} \right.
\eeq{Veom}
The boundary term from varying $Y$ reads\footnote{Note that there are
no boundary terms arising from the variation with respect to $V_\pm$,
nor are there any $v_\pm$-terms in the boundary term that might contribute,
since we are using the covariant variables $\widehat\psi^\mu_\pm$
rather than the combination $\psi^\mu_\pm + k^\mu v_\pm$.}
\beq
\int d\tau \,\, \left[
 -(v_{++}+v_{--}) \delta y + v_+ \delta y_+ + v_- \delta y_- \right] =0\,,
\eeq{ybeom}
where
$$
v_{\pm\pm} \equiv (D_{\pm} V_{\pm})|_{\theta=0} \,,
\quad\quad
y_{\pm}  \equiv  (D_{\pm} Y)|_{\theta=0}\, , 
\quad\quad
y  \equiv Y|_{\theta=0} \,.
$$
In the interest of generality, we need to be careful when solving
(\ref{ybeom}). Since $Y$ is a Lagrange multiplier, so we know that
it is definitely independent of 
all other fields, it is safest -- and easiest --
to first analyse the components of this field rather than
those of $V_\pm$. More precisely, we need to find the most general
boundary condition relating $y_+$ to $y_-$, and then substitute
this relation in (\ref{ybeom}) before solving it.
Dimensional considerations (classical conformality),
analogous to those that led to the most
general linear fermionic ansatz in the original model, suggest that
the ansatz for $y_\pm$ must take the form
\beq
y_- + a y_+ =0\,,
\eeq{yans1}
where $a$ is some scalar field. Because $Y$ is independent of the other
fields, $a$ must also be independent of them. However, $a$ may a priori
depend on $y$. We can determine whether or not this is the case by
examining the superpartner of (\ref{yans1}), which reads
$$
a\d_\+ y + \eta \d_= y = (\eta a -1) y_{+-}
-\d_y a (y_+ + \eta y_-) y_+ \,.
$$
Due to (\ref{yans1}), the last term on the right-hand side is
annihilated by $y_+$, and we have
\beq
a\d_\+ y + \eta \d_= y = (\eta a -1) y_{+-} \,.
\eeq{yans3}
On the other hand
we know from Section~\ref{bosbulk} that (\ref{yans3}) should
reduce to $\d_\tau y=0$ in the bosonic case. Since 
in the bosonic model $y_{+-}=0$, this is true if and only if $a=\eta$.
Hence the most general ansatz for $y_\pm$ is
\beq
y_- + \eta y_+ =0\,.
\eeq{yans}

Returning to the equation of motion (\ref{ybeom}), the insertion
of (\ref{yans}) reduces it to
\beq
 -(v_{++}+v_{--}) \delta y + (v_+ -\eta v_-) \delta y_+ =0\,.
\eeq{ybeom3}
The two terms must vanish independently.\footnote{There were at the outset
the three independent fields $\delta y, \delta y_\pm$, and we have used
(\ref{yans}) to eliminate $\delta y_-$, leaving the two independent
fields $\delta y, \delta y_+$.} From the first term it is clear that
$\delta y=0$, because $v_{++}+v_{--}$ is not allowed to vanish, for
the same reason (gauge freedom) that $A_{\+} + A_{=}$ had to be nonzero in
Section~\ref{bosbulk}. More precisely, the gauge transformation
of $v_{++} + v_{--}=0$ is $-2i \d_\tau \epsilon=0$, which would restrict
the gauge. We thus find
$$
\d_\tau y = 0 \, . \hspace{3cm} (\ref{dty0})
$$
The second term in (\ref{ybeom3}) allows two possibilities:
\beq
v_- -\eta v_+ =0
\eeq{ysoln1}
or
\beq
\delta y_+ =0 \,.
\eeq{ysoln2}
However, it turns out that the latter condition is not a good solution,
because it does not eliminate $y_+$ on the boundary when used in
the parent action (which was the whole point of using
the $Y$ field equations in the action). To see this, consider the
boundary $Y$-terms remaining after substituting the bulk equations
of motion (\ref{Yeom2}) in (\ref{gaugeact1}),
$$
 \int d\tau \left[ -2\lambda \d_\tau y +i (v_+ -\eta v_-) y_+ \right ]\,,
$$
where $\lambda \equiv \Lambda|_{\theta=0}$.
The first term vanishes due to (\ref{dty0}),
and (\ref{ysoln1}) cancels the second term, whereas (\ref{ysoln2})
would not cancel it. Hence we must adopt (\ref{ysoln1}), and
this is the boundary equation of motion for $y_+$.

In Table~\ref{Yonshell} we summarise the boundary conditions implied
by $Y$-on-shell analysis.  Note that (\ref{dty0}) and (\ref{yans})
are off-shell conditions imposed by hand, whereas (\ref{ysoln1}) a
priori holds only on-shell.  However, we will see presently that
in fact this condition also must be off-shell.

\begin{table}[ht]
\vspace{1cm}
\begin{tabular*}{\textwidth}{@{\extracolsep{\fill}}|ccc|}\hline
& $
\begin{array}{rcl@{\hspace{2cm}}l}
\d_\tau y      &=& 0 & (\ref{dty0})   \\
y_- + \eta y_+ &=& 0 & (\ref{yans})    \\
v_- -\eta v_+  &=& 0 & (\ref{ysoln1})
\end{array}
$ & \\ \hline
\end{tabular*}
\caption{Conditions found when integrating out the $Y$-field
in the supersymmetric parent action (\ref{gaugeact1}).}
\vspace{1cm}
\label{Yonshell}
\end{table}

\subsubsection{Recovering the original model}
\label{recorig}

Having derived the most general field equations for $Y$,
we now use them to integrate out $Y$ in the parent action
(\ref{gaugeact1}). Thus, using Eq.~(\ref{Yeom2}) and the
conditions in Table~\ref{Yonshell}, the result is
\ber
\nonumber  S_\Lambda &=& \int d^2\xi d^2\theta\,\,\,
\left [(D_+ \Phi^\mu+k^\mu D_+\Lambda)
 ( D_- \Phi^\nu + k^\nu D_-\Lambda)E_{\mu\nu}
  +\omega_\mu D_{(+} \Phi^\mu D_{-)} \Lambda \right ] \\
&& +\frac{i}{2}\int d\tau\,\,  B_{\mu\nu} [\widehat\psi_{+}^\mu
\widehat\psi_{+}^\nu + \widehat\psi_-^\mu \widehat\psi_-^\nu ] \, .
\eer{susynoY}
This action must be gauge invariant. Its gauge transformation,
using (\ref{Lg0}) and (\ref{domega}), is a pure boundary term,
\ber
\nonumber \delta_k S_\Lambda &=& \int d\tau \,\,\,
\left [2\epsilon \,\, \omega_\mu (\d_\tau X^\mu +k^\mu \d_\tau \lambda)
+i \epsilon \,\, (\omega_{\nu,\mu} k^\mu + \omega_\mu k^\mu_{\,\,,\nu})
(\lambda_+ \widehat\psi_{+}^\nu + \lambda_- \widehat\psi_{-}^\nu) \right. \\
&& \left . -i \epsilon_+ \,\, \omega_\mu \widehat\psi_{+}^\mu
-i \epsilon_- \,\, \omega_\mu \widehat\psi_{-}^\mu \right] \, ,
\eer{gvarnoY}
where
$$
\epsilon_{\pm} \equiv (D_{\pm} \epsilon)|_{\theta=0} \,,
\quad\quad
\lambda_\pm \equiv (D_{\pm} \Lambda)|_{\theta=0} \,.
$$
Note that the first term in (\ref{gvarnoY}) is 
the expected bosonic part of the variation, cf.~Eq.~(\ref{bosgnoy}).
To find the conditions necessary for $\delta_k S_\Lambda$ to vanish
we make use of the fact that gauge invariance
of (\ref{ysoln1}) yields a relation between the gauge parameters
$\epsilon_\pm$,
\beq
\epsilon_- -\eta \epsilon_+ =0 \, .
\eeq{epsrel}
In addition, (\ref{ysoln1}) implies
$\lambda_- -\eta \lambda_+ =0$, which we insert together with
(\ref{epsrel}) in (\ref{gvarnoY}) to get
\ber
\nonumber \delta_k S_\Lambda &=& \int d\tau \,\,\,
\left [2\epsilon \,\, \omega_\mu (\d_\tau X^\mu +k^\mu \d_\tau \lambda)
+i \epsilon \,\, (\omega_{\nu,\mu} k^\mu + \omega_\mu k^\mu_{\,\,,\nu})
v_+ (\widehat\psi_{+}^\nu + \eta \widehat\psi_{-}^\nu) \right. \\
&& \left .
-i \epsilon_+ \,\, \omega_\mu (\widehat\psi_{+}^\mu
                                +\eta \widehat\psi_{-}^\mu) \right] \, .
\eer{gvarnoY1}
The vanishing of the first term implies, using the Dirichlet condition
(\ref{QdX}),
$$
\pi^\mu_{\,\,\,\nu} \omega_\mu = 0 \,, \hspace{3cm} (\ref{piomega0})
$$
which also annihilates the last term because, due to the fermionic
Dirichlet condition (\ref{Qpsi}) and $Q^\mu_{\,\,\,\nu} k^\nu =0$,
Eq.~(\ref{Qk0}), $\widehat\psi^\mu_\pm$ satisfy
\beq
Q^\mu_{\,\,\,\nu}
 \left( \widehat\psi^\nu_+ + \eta \widehat\psi^\nu_- \right) =0\, .
\eeq{Qpsi2}
Because (\ref{piomega0}) implies $k^\mu \omega_\mu=0$
and therefore $(k^\mu \omega_\mu)_{,\nu}=0$,
the remaining term in (\ref{gvarnoY}) may be rewritten as
$$
i \epsilon \,\, v_+ k^\lambda \pi^\mu_{\,\,\,\lambda}
 \omega_{[\rho,\mu]} \pi^\rho_{\,\,\,\nu}
(\widehat\psi_{+}^\nu + \eta \widehat\psi_{-}^\nu) \,,
$$
which implies
\beq
k^\lambda \pi^\mu_{\,\,\,\lambda} \left( \L_k B_{\mu\rho} \right)
\pi^\rho_{\,\,\,\nu} =0\, ,
\eeq{kLkpiE}
i.e., Eq.~(\ref{LkpiE}) contracted with $k^\mu$.

The action (\ref{susynoY}) is thus gauge invariant provided
(\ref{piomega0}), (\ref{epsrel}) and (\ref{kLkpiE})
are satisfied on the boundary, and we can now go to adapted
coordinates ($k^\mu = \delta^\mu_0$), and fix the gauge such that
$\Lambda = 0$. The result is the original sigma model action,
$$
\begin{array}{rcl@{\hspace{2cm}}l}
S &=& \int d^2\xi d^2\theta \,\,\,
D_+ \Phi^\mu D_- \Phi^\nu E_{\mu\nu} 
+\frac{i}{2}\int d\tau\,\,  B_{\mu\nu} [\psi_{+}^\mu
\psi_{+}^\nu + \psi_-^\mu \psi_-^\nu ]\, .   & (\ref{goodaction})
\end{array}
$$

In Table~\ref{Trec} we collect the conditions necessary for
a consistent retrieval of the original supersymmetric sigma model
from the parent action (\ref{gaugeact1}), assuming the conditions
in Table~\ref{Yonshell} are satisfied.

\begin{table}[ht]
\vspace{1cm}
\begin{tabular*}{\textwidth}{@{\extracolsep{\fill}}|ccc|}\hline
& $
\begin{array}{rcl@{\hspace{2cm}}l}
Q^\mu_{\,\,\,\nu}\d_\tau X^\nu &=& 0 &             (\ref{QdX}) \\
Q^\mu_{\,\,\,\nu} k^\nu &=&0 &                     (\ref{Qk0}) \\
\L_{k} g_{\mu\nu} &=&0 &                           (\ref{Lg0}) \\
\L_k B_{\mu\nu} - \omega_{[\nu,\mu]}&=&0 &     (\ref{domega}) \\
\pi^\mu_{\,\,\,\nu} \omega_\mu &=& 0 &        (\ref{piomega0}) \\
\epsilon_- -\eta \epsilon_+  &=& 0 &            (\ref{epsrel}) \\
Q^\mu_{\,\,\,\nu} \left( \widehat\psi^\nu_+
    + \eta \widehat\psi^\nu_- \right)  &=& 0 &   (\ref{Qpsi2}) \\
k^\lambda \pi^\mu_{\,\,\,\lambda} \left( \L_k B_{\mu\rho} \right)
\pi^\rho_{\,\,\,\nu} &=& 0 &                    (\ref{kLkpiE})
\end{array} 
$ & \\ \hline
\end{tabular*}
\caption{Conditions necessary for
a consistent retrieval of the original supersymmetric sigma model
from the parent action (\ref{gaugeact1}). We have assumed that the conditions
in Table~\ref{Yonshell} are satisfied.}
\vspace{1cm}
\label{Trec}
\end{table}

\subsubsection{Finding the dual model}
\label{finddual}

To find the dual model we use the bulk equations of motion
(\ref{Veom}) to integrate out $V_\pm$
in (\ref{gaugeact1}), resulting in the following action,
\ber
\nonumber  S_Y &=& \int d^2\xi d^2\theta \,\,\,
\left [ \high D_+ \Phi^\mu  D_- \Phi^\nu E_{\mu\nu} \right . \\
\nonumber  &+& \left. (k^\lambda k^\sigma E_{\lambda\sigma})^{-1} \left(D_+ Y +
\omega_\mu D_+ \Phi^\mu + k^\mu E_{\nu\mu} D_+ \Phi^\nu \right)
\left(D_- Y + \omega_\rho D_- \Phi^\rho - k^\gamma E_{\gamma\rho} D_- \Phi^\rho \right)
 \high \right ]  \\
& +&\frac{i}{2}\int d\tau\,\,  B_{\mu\nu} [\widehat\psi_{+}^\mu
\widehat\psi_{+}^\nu + \widehat\psi_-^\mu \widehat\psi_-^\nu ] \,.
\eer{susynoA}
This action should be gauge invariant, and its
gauge transformation is a pure boundary term,
\ber
\nonumber \delta_k S_Y &=& \int d\tau \,\,\,
\left [2\epsilon ( \d_\tau y + \omega_\mu \d_\tau X^\mu)
+ i k^\mu \omega_\mu (\epsilon_+ v_+ + \epsilon_-v_-)
+i \epsilon k^\nu \omega_{[\nu,\mu]}
(\widehat\psi_{+}^\mu v_+ +  \widehat\psi_{-}^\mu v_-)  \right . \\
&& \left.
-i \omega_\mu (\epsilon_+ \widehat\psi_{+}^\mu
+ \epsilon_- \widehat\psi_{-}^\mu)
-i ( \epsilon_+ y_+ + \epsilon_- y_-)
 \right] \,,
\eer{gvarnoV}
which vanishes when we use the conditions listed in Tables~\ref{Yonshell}
and~\ref{Trec}, showing that $S_Y$ is gauge invariant. Note
how this requires (\ref{ysoln1}) to be an off-shell condition.

Going to adapted coordinates and fixing the gauge such that $\Phi^0=0$ (and
hence $X^0=\psi_{\pm}^0=0$), we obtain the dual action,
\ber
\nonumber  \widetilde{S}' &=& \int d^2\xi d^2\theta\,\,\,
\left [D_+ Y D_- Y \widetilde{E}_{00}
+ D_+ Y D_- \Phi^n \widetilde{E}_{0n} 
+ D_+ \Phi^m D_- Y \widetilde{E}_{m0}
+ D_+ \Phi^m D_- \Phi^n \widetilde{E}_{mn} \right] \\
 &+& \frac{i}{2}\int d\tau\,\, \left\{
E_{mn} [\widehat\psi_{+}^m \widehat\psi_{+}^n
 + \widehat\psi_-^m \widehat\psi_-^n ]
+ E_{0n} [\widehat\psi_{+}^0 \widehat\psi_{+}^n
 + \widehat\psi_-^0 \widehat\psi_-^n ]
+ E_{n0} [\widehat\psi_{+}^n \widehat\psi_{+}^0
 + \widehat\psi_-^n \widehat\psi_-^0 ]
\right\} \, ,
\eer{susydual1}
where $\widetilde{E}_{\mu\nu}$ is the dual background given by
Buscher's rules (\ref{Buscher}), except now it is a superfield.
To rewrite the boundary term in $\widetilde{S}'$
in terms of dual quantities, we first observe that, in the chosen gauge,
\beq
\widehat\psi_{\pm}^0 \equiv \psi_{\pm}^0 + v_\pm = v_\pm \, , \quad\quad
\widehat\psi_{\pm}^n \equiv \psi_{\pm}^n \, .
\eeq{hatpsiv}
We moreover rewrite the component form of the
$V_\pm$ bulk equations of motion (\ref{Veom})
in terms of the dual background,
\beq
\left\{
\begin{array}{l}
v_+ \widetilde E_{00}^{-1}
+ \widetilde E_{00}^{-1} \widetilde E_{n0} \psi^n_+ + y_+ = 0\,, \\
v_- \widetilde E_{00}^{-1}
- \widetilde E_{00}^{-1} \widetilde E_{0n} \psi^n_- - y_- = 0 \,.
\end{array} \right.
\eeq{veom1}
Given $y_- + \eta y_+=0$, Eq.~(\ref{yans}),
and $v_- - \eta v_+=0$, Eq.~(\ref{ysoln1}),
the first equation minus $\eta\equiv\pm$1 times the second, and
restricting to the boundary, yields the relation
\beq
\widetilde E_{n0} \psi^n_+ = - \eta \widetilde E_{0n} \psi^n_- \,.
\eeq{veom2}
Using (\ref{hatpsiv}) and (\ref{veom2}) as well as (\ref{piomega0}) and
Buscher's rules (\ref{Buscher}), we may now rewrite the dual action
(\ref{susydual1}) as
\beq
\widetilde{S}= \int d^2\xi d^2\theta\,\,\,
D_+ \widetilde \Phi^\mu D_- \widetilde \Phi^\nu \widetilde{E}_{\mu\nu}
+ \frac{i}{2}\int d\tau\,\,
\widetilde B_{\mu\nu} [\widetilde \psi_{+}^\mu \widetilde \psi_{+}^\nu
 + \widetilde \psi_-^\mu \widetilde \psi_-^\nu ] \,,
\eeq{susydual2}
where
$$
\widetilde \Phi^0 \equiv Y \,,\quad\quad
\widetilde \Phi^n \equiv \Phi^n \,, \quad\quad
\widetilde \psi_{\pm}^0 \equiv y_{\pm} \,, \quad\quad
\widetilde \psi_{\pm}^n \equiv \psi_{\pm}^n \,.
$$
Thus the dual action has the same form as the original action, with
$Y$ playing the role of a $\Phi^0$-coordinate in the dual coordinate
system, and $y_\pm$ being the dual analogues of $\psi^0_\pm$.

\subsection{Gauge invariance of parent action}
\label{actioninv}

It is easily verified that the gauge transformation of the parent
action before integrating out the fields $V_\pm$ is the same as after
integrating them out, i.e., Eq.~(\ref{gvarnoV}). As already discussed,
this vanishes when all the boundary conditions listed in
Tables~\ref{Yonshell} and~\ref{Trec} are satisfied, hence the parent
action (\ref{gaugeact1}) is indeed gauge invariant.

\subsection{Compatibility of field equations, part I}
\label{eomI}

Now we want to compare the boundary equations of motion obtained from
the parent action with the bulk equations restricted to the boundary,
to check if they are compatible. The boundary field equations for $Y$
were derived in Section~\ref{susybulk} and are listed in
Table~\ref{Yonshell}.  There are no boundary equations of motion for
$V_\pm$, so we move on to derive the $X^\mu$ and $\widehat
\psi^\mu_\pm$ field equations. The boundary term that results from
varying the parent action (\ref{gaugeact1}) with respect to $X^\mu$
and $\widehat \psi^\mu_\pm$ in component form reads
\ber
\nonumber \delta \widehat S
 &=& \int d\tau  \,\, \left\{ \delta X^\rho \,\, \left[ \high
\d_= X^\mu E_{\rho\mu} - \d_\+ X^\mu E_{\mu\rho}  \right. \right.\\
\nonumber  && \left.\left.
+i \widehat \psi^\mu_+ v_+ ( \L_k B_{\mu\rho}- \omega_{\rho,\mu})
-i \widehat \psi^\mu_- v_- ( \L_k B_{\rho\mu}+ \omega_{\rho,\mu})
\right. \right. \\
\nonumber  && \left. \left.
- i(v_{++}+v_{--})\omega_\rho
+ i v_{++}k^\nu E_{\nu\rho}  - i v_{--}k^\nu E_{\rho\nu} 
\right. \right. \\
\nonumber  && \left. \left.
+i \widehat \psi^\mu_+\widehat \psi^\nu_+
 \left( E_{\nu\rho,\mu}+\half B_{\mu\nu,\rho} \right)
+i \widehat \psi^\mu_-\widehat \psi^\nu_-
 \left( E_{\rho\mu,\nu}+\half B_{\mu\nu,\rho} \right)
\right] \right. \\
 && \left.
+i \delta \widehat \psi^\mu_+
 \left[ \widehat \psi^\nu_+ g_{\mu\nu} - v_+ \omega_\mu \right]
-i \delta \widehat \psi^\mu_-
 \left[ \widehat \psi^\nu_- g_{\mu\nu} + v_- \omega_\mu  \right]
\high \right\} \, .
\eer{susXbeom}
Using $\delta X^\rho = \delta X^\lambda \pi^\rho_{\,\,\,\lambda}$
as well as (\ref{Lg0}), (\ref{domega}) and (\ref{piomega0}),
$\delta \widehat S$ reduces to
\ber
\nonumber \delta \widehat S &=& \int d\tau  \,\,
\left\{\delta X^\lambda \pi^\rho_{\,\,\,\lambda} \,\, \left[ \high
\d_= X^\mu E_{\rho\mu} - \d_\+ X^\mu E_{\mu\rho}  \right. \right.\\
\nonumber  && \left.\left.
-i \left( \widehat \psi^\mu_+ v_+ + \widehat \psi^\mu_- v_-\right)
 \omega_{\mu,\rho}
+ i v_{++}k^\nu E_{\nu\rho}  - i v_{--}k^\nu E_{\rho\nu} 
\right. \right. \\
\nonumber  && \left. \left.
+i \widehat \psi^\mu_+\widehat \psi^\nu_+
 \left( E_{\nu\rho,\mu}+\half B_{\mu\nu,\rho} \right)
+i \widehat \psi^\mu_-\widehat \psi^\nu_-
 \left( E_{\rho\mu,\nu}+\half B_{\mu\nu,\rho} \right)
\right] \right. \\
 && \left.
+i \delta \widehat \psi^\mu_+
 \left[ \widehat \psi^\nu_+ g_{\mu\nu} - v_+ \omega_\mu \right]
-i \delta \widehat \psi^\mu_-
 \left[ \widehat \psi^\nu_- g_{\mu\nu} + v_- \omega_\mu  \right]
\high \right\} \, .
\eer{susXbeom2}
We will return to this equation to derive the most general equation
of motion, but first we turn to the $V_\pm$ bulk equations of motion,
Eq.~(\ref{Veom}), restricting their component form to the boundary.
Taking the first equation minus $\eta$ times the second one, we obtain
a superfield relation whose fermionic lowest component reads
\beq
0= y_+ + \eta y_- - (v_+ + \eta v_-) k^\mu \omega_\mu
+      \widehat \psi^\mu_+ \left( \omega_\mu + k^\nu E_{\mu\nu} \right)
+ \eta \widehat \psi^\mu_- \left( \omega_\mu - k^\nu E_{\nu\mu} \right) \,  ,
\eeq{Veomcomb1}
and its bosonic superpartner reads
\ber
\nonumber 0 &=& -2\d_\tau y  - 2\omega_\mu \d_\tau X^\mu
-\d_\+ X^\mu  k^\nu E_{\mu\nu} +\d_= X^\mu   k^\nu E_{\nu\mu}
+i(v_{++} - v_{--})k^\mu k^\nu g_{\mu\nu} \\
&&
+i\widehat \psi^\rho_+\widehat \psi^\nu_+
 (\omega_{\nu} + k^\mu E_{\nu\mu})_{,\rho}
+i\widehat \psi^\rho_-\widehat \psi^\nu_-
 (\omega_{\nu} - k^\mu E_{\mu\nu})_{,\rho} \, .
\eer{Veomcomb2}
Inserting (\ref{dty0}), (\ref{piomega0}), (\ref{yans}) and the Dirichlet
condition (\ref{Qpsi2}), Eqs.~(\ref{Veomcomb1})
and (\ref{Veomcomb2}) simplify to
\beq
0= k^\nu E_{\mu\nu} \widehat \psi^\mu_+
 - \eta k^\nu E_{\nu\mu} \widehat \psi^\mu_-  \, ,
\eeq{Veomcomb3}
\ber
\nonumber 0 &=& 
k^\nu \left( \d_= X^\mu E_{\nu\mu} - \d_\+ X^\mu E_{\mu\nu}  \right) 
+i(v_{++} - v_{--})k^\mu k^\nu g_{\mu\nu} \\
&&
+i\widehat \psi^\rho_+\widehat \psi^\nu_+
 (\omega_{\nu} + k^\mu E_{\nu\mu})_{,\rho}
+i\widehat \psi^\rho_-\widehat \psi^\nu_-
 (\omega_{\nu} - k^\mu E_{\mu\nu})_{,\rho} \, .
\eer{Veomcomb4}
The $\widehat\psi\widehat\psi$-terms can be rewritten
using ${\cal L}_k B_{\mu\nu} = \omega_{[\nu,\mu]}$ so that
(\ref{Veomcomb4}) becomes
\ber
\nonumber 0 &=& 
k^\nu \left( \d_= X^\mu E_{\nu\mu} - \d_\+ X^\mu E_{\mu\nu}  \right) 
+i(v_{++} - v_{--})k^\mu k^\nu g_{\mu\nu} \\
\nonumber &&
+ i\widehat \psi^\mu_+\widehat \psi^\nu_+
 \left(E_{\nu\rho,\mu}+\half B_{\mu\nu,\rho}\right)k^\rho 
+i\widehat \psi^\mu_-\widehat \psi^\nu_-
 \left(E_{\rho\mu,\nu}+\half B_{\mu\nu,\rho}\right)k^\rho \\
&& +i(\widehat \psi^\mu_+\widehat \psi^\nu_+
 - \widehat \psi^\mu_-\widehat \psi^\nu_-) k^\rho_{,\mu} g_{\rho\nu} \, .
\eer{Veomcomb5}

This is as far as we get without employing a boundary ansatz for the
fermions.  Otherwise we cannot derive the most general equations of
motion from (\ref{susXbeom2}), to see if they are compatible with
(\ref{Veomcomb5}).  We therefore turn to the fermionic ansatz next,
deducing the appropriate one in a fashion similar to the bosonic
analysis in Section~\ref{bosansatz}.

\subsection{Intermission: boundary ansatz}
\label{intermission}

For the original nonlinear sigma model (\ref{goodaction}) we found that
the appropriate fermionic ansatz was (\ref{fermans}), accompanied by
certain restrictions on the gluing matrix $R^\mu_{\,\,\,\nu}$ dictated
by superconformal symmetry.  Here it is natural to
make the analogous ansatz for the gauged model,
\beq
\widehat \psi^\mu_- = \eta R^\mu_{\,\,\,\nu}  \widehat \psi^\nu_+  \,.
\eeq{fermans3}
The form of this ansatz is also suggested by the supersymmetric generalisation
of the bosonic ansatz (\ref{bosgans}).

In analogy with the bosonic analysis in Section~\ref{bosansatz}, we
may deduce that the gluing matrix here is the same as in the ungauged
model. The currents associated to superconformal invariance are
identical in form to those of the original model, as shown in
Appendix~\ref{gaugecur}. Thus the superconformal conditions are also
of the same form, and in the fixed gauge they should reduce to the
conditions of the original model. But again there are no consistent
conformally allowed modifications to $R^\mu_{\,\,\,\nu}$.

It is important for the continued analysis to realise that because the
analysis of superconformal currents is identical to that of the
original model, we know that all the boundary conditions in
Table~\ref{N1bc} must hold also in the gauged model. In particular,
$\pi$-integrability together with $\pi^\mu_{\,\,\,\nu} \omega_\mu=0$
implies $\pi^\mu_{\,\,\,\sigma} (\L_k B_{\mu\nu})
\pi^\nu_{\,\,\,\rho}=0$, Eq.~(\ref{LkpiE}), hence (\ref{kLkpiE}) is
automatically satisfied. Therefore we may henceforth replace
(\ref{kLkpiE}) with the condition (\ref{LkpiE}).

Just like in Section~\ref{1Dformalism} we can write (\ref{fermans3})
and its bosonic superpartner in terms of 1D superfields. The
supersymmetry transformation of the fermionic ansatz reads
\ber
\nonumber  i\d_= X^\mu &=&
i R^\mu_{\,\,\,\nu} \d_\+ X^\nu - 2 \eta P^\mu_{\,\,\,\nu} F^\nu_{+-}
+ 2R^\mu_{\,\,\,\nu,\rho} P^\rho_{\,\,\,\lambda}
     \widehat\psi^\lambda_+ \widehat\psi^\nu_+ \\
\nonumber &&
- R^\mu_{\,\,\,\nu,\rho} k^\rho v_+ \widehat\psi^\nu_+
+ 2v_+ \left( - R^\mu_{\,\,\,\nu} k^\nu_{\,\,,\rho} + k^\mu_{\,\,,\rho} \right)
          P^\rho_{\,\,\,\lambda} \widehat\psi^\lambda_+ \\
&& + \left( R^\mu_{\,\,\,\nu} k^\nu - k^\mu \right)
  \left( v_{++} + \eta v_{-+}\right) \, ,
\eer{dmXans}
where we have used (\ref{fermans3}) as well as the boundary relation
for $v_\pm$, Eq.~(\ref{ysoln1}), and its superpartner condition,
\beq
v_{--} + \eta v_{+-} = v_{++} + \eta v_{-+} \, .
\eeq{vpp}
In terms of 1D superfields, (\ref{fermans3}) and (\ref{dmXans}) are
summarised in the relation
\beq
(\nabla K^\mu + \widehat S^\mu) = R^\mu_{\,\,\,\nu} (K)
  (\nabla K^\nu -\widehat S^\nu) \, ,
\eeq{hatKSbc}
where
$$
\nabla K^\mu \equiv D K^\mu + k^\mu V_1 \, ,
\quad\quad \widehat S^\mu \equiv S^\mu + k^\mu V_2 \, ,
$$
with
$$
V_1\equiv \half (V_- + \eta V_+)\, ,
\quad\quad  V_2 \equiv \half (V_- - \eta V_+) \,.
$$
It immediately follows that
the fermionic Dirichlet condition (\ref{Qpsi2})
and its superpartner (\ref{QdX}) may be written analogously to (\ref{QDK}),
\beq
Q^\mu_{\,\,\,\nu} \nabla K^\nu   =0 \, .
\eeq{QnablaK}

It is now easy to check that gauge invariance of the ansatz (\ref{hatKSbc})
implies that the Lie derivative of $R^\mu_{\,\,\,\nu}$ must vanish,
$$
\begin{array}{rcl@{\hspace{3.6cm}}l}
{\cal L}_k R^\mu_{\,\,\,\nu} &=&0 \, .& (\ref{LieR})
\end{array}
$$
This is very useful in the continued analysis, and it also leads to
$$
\begin{array}{rcl@{\hspace{2cm}}l}
{\cal L}_k \pi^\mu_{\,\,\,\nu}
 = {\cal L}_k Q^\mu_{\,\,\,\nu} &=&0 \, .& (\ref{LieQpi})
\end{array}
$$

\subsection{Compatibility of field equations, part II}
\label{eomII}

Let us now use the ansatz (\ref{fermans3}) in the $V_\pm$ field
equations (\ref{Veomcomb3}) and (\ref{Veomcomb5}).  First we note that
(\ref{Veomcomb3}) is automatically satisfied by virtue of the
superconformality conditions (\ref{pigQ}) and (\ref{piEpiR}).  Next we
use (\ref{RgRcond1}), (\ref{LieR}) and (\ref{fermans3}) to rewrite
the last term in (\ref{Veomcomb5}). The $V_\pm$ field equation then
becomes
\ber
\nonumber 0 &=& 
k^\nu \left( \d_= X^\mu E_{\nu\mu} - \d_\+ X^\mu E_{\mu\nu}  \right) 
+i(v_{++} - v_{--})k^\mu k^\nu g_{\mu\nu} \\
\nonumber &&
+ i\widehat \psi^\mu_+\widehat \psi^\nu_+
 \left(E_{\nu\rho,\mu}+\half B_{\mu\nu,\rho}\right)k^\rho 
+i\widehat \psi^\mu_-\widehat \psi^\nu_-
 \left(E_{\rho\mu,\nu}+\half B_{\mu\nu,\rho}\right)k^\rho \\
&& -i\widehat \psi^\mu_+ \widehat \psi^\nu_+
R^\gamma_{\,\,\,\nu} g_{\gamma\rho} R^\rho_{\,\,\,\mu,\lambda} k^\lambda \,.
\eer{Veomcomb7}
We will do the same for the $X^\mu$ and $\widehat\psi^\mu_\pm$ field
equation (\ref{susXbeom2}), but first we need the field
variation of the fermionic ansatz,
\beq
\delta \widehat \psi^\mu_-
 = \eta R^\mu_{\,\,\,\nu} \delta \widehat \psi^\nu_+
+ \eta R^\mu_{\,\,\,\nu,\rho} \widehat \psi^\nu_+ \delta X^\rho .
\eeq{fermansvar1}
We use this to substitute for $\delta \widehat \psi^\mu_-$
in (\ref{susXbeom2}), and we moreover use (\ref{RgRcond1}) 
as well as (\ref{ysoln1}) and (\ref{fermans3})
in the $\delta \widehat \psi^\mu_+$-terms.
Eq.~(\ref{susXbeom2}) then becomes
\ber
\nonumber \delta \widehat S &=& \int d\tau  \,\,
\left\{\delta X^\lambda \pi^\rho_{\,\,\,\lambda} \,\, \left[ \high
\d_= X^\mu E_{\rho\mu} - \d_\+ X^\mu E_{\mu\rho}
+ i v_{++} k^\nu E_{\nu\rho}  - i v_{--}k^\nu E_{\rho\nu} 
 \right. \right.\\
\nonumber  && \left.\left.
+i \widehat \psi^\mu_+\widehat \psi^\nu_+
 \left( E_{\nu\rho,\mu}+\half B_{\mu\nu,\rho} \right)
+i \widehat \psi^\mu_-\widehat \psi^\nu_-
 \left( E_{\rho\mu,\nu}+\half B_{\mu\nu,\rho} \right)
\right. \right. \\
&& \left. \left.
-i \widehat \psi^\mu_+ \widehat \psi^\nu_+
R^\gamma_{\,\,\,\nu} g_{\gamma\lambda} R^\lambda_{\,\,\,\mu,\rho}
-2i \widehat \psi^\mu_+ v_+ \left(P^\nu_{\,\,\,\mu} \omega_\nu \right)_{,\rho}
\right] 
-2i\, \delta \widehat \psi^\mu_+ \, v_+ P^\nu_{\,\,\,\mu} \omega_\nu 
\high \right\} \, .
\eer{susXbeom3}
The last term in the square bracket as well as the
$\delta \widehat \psi^\mu_+$-term vanish by virtue of
(\ref{piomega0}) since $P^\nu_{\,\,\,\mu} \omega_\nu =
P^\nu_{\,\,\,\mu} \pi^\rho_{\,\,\,\nu} \omega_\rho=0$. 
Thus the final, most general, $X^\mu$ boundary equation of motion reads
\ber
\nonumber  0 &=& \pi^\rho_{\,\,\,\lambda} \,\, \left[
\d_= X^\mu E_{\rho\mu} - \d_\+ X^\mu E_{\mu\rho}
+ i v_{++} k^\nu E_{\nu\rho}  - i v_{--}k^\nu E_{\rho\nu} 
 \right. \\
\nonumber  && \left.
+i \widehat \psi^\mu_+\widehat \psi^\nu_+
 \left( E_{\nu\rho,\mu}+\half B_{\mu\nu,\rho} \right)
+i \widehat \psi^\mu_-\widehat \psi^\nu_-
 \left( E_{\rho\mu,\nu}+\half B_{\mu\nu,\rho} \right)
\right.  \\
&&  \left.
-i \widehat \psi^\mu_+ \widehat \psi^\nu_+
R^\gamma_{\,\,\,\nu} g_{\gamma\sigma} R^\sigma_{\,\,\,\mu,\rho}
\right] .
\eer{susXbeom5}
It is immediately clear that (\ref{susXbeom5}) contracted with $k^\lambda$ is
precisely Eq.~(\ref{Veomcomb7}), hence the $X^\mu$ field equations
are compatible with the $V_\pm$ equations on the boundary, which is
what we wanted to show.

\subsection{Gauge invariance of boundary conditions}
\label{gauging}

Next we check gauge invariance of all the boundary conditions used in
this section.  We start with the superconformal conditions in
Table~\ref{N1bc}, and then turn to the conditions found in
Sections~\ref{susybulk}--\ref{eomII}.  The only condition in
Table~\ref{N1bc} that is not trivially gauge invariant is the
Dirichlet condition (\ref{QDK}).  Under a gauge transformation it
becomes, using
$Q^\mu_{\,\,\,\nu}k^\nu= Q^\mu_{\,\,\,\nu}DK^\nu = 0$,
$$
k^\sigma \pi^\rho_{\,\,\,\sigma} Q^\mu_{\,\,\,[\nu,\rho]}
\pi^\nu_{\,\,\,\lambda} DK^\lambda  =0\,,
$$
which is satisfied due to $\pi$-integrability, Eq.~(\ref{piinteg1}).
The analysis of the corresponding condition
(\ref{QnablaK}) for covariant fields is identical, and we already checked in
Section~\ref{intermission} that the ansatz (\ref{hatKSbc}) is gauge invariant.
It remains to check the following conditions:
$$
\begin{array}{rcl@{\hspace{2cm}}l}
Q^\mu_{\,\,\,\nu} k^\nu&=&0 &                               (\ref{Qk0}) \\
{\cal L}_{k} g_{\mu\nu} &=&0 &                              (\ref{Lg0}) \\
{\cal L}_{k} B_{\mu\nu} -\d_{[\mu} \omega_{\nu]}&=&0 &   (\ref{domega}) \\
{\cal L}_k R^\mu_{\,\,\,\nu} &=&0  &                       (\ref{LieR}) \\
{\cal L}_k \pi^\mu_{\,\,\,\nu} =
 {\cal L}_k Q^\mu_{\,\,\,\nu} &=&0  &                    (\ref{LieQpi}) \\
\pi^\mu_{\,\,\,\lambda} \left( \L_k B_{\mu\rho} \right)
\pi^\rho_{\,\,\,\nu} &=& 0  &                             (\ref{LkpiE}) \\
\d_\tau y &=&0   &                                         (\ref{dty0}) \\
\pi^\mu_{\,\,\,\nu} \omega_\mu &=&0   &                (\ref{piomega0}) \\
D_+V_- + D_-V_+ &=&0  &                                    (\ref{Yeom}) \\
V_+ k^\mu k^\nu E_{\mu\nu}+ D_+ \Phi^\nu \left( k^\mu E_{\nu\mu}
    + \omega_\nu \right) + D_+ Y &=&0  &         (\ref{Veom})\mathrm{a} \\
V_- k^\mu k^\nu E_{\mu\nu} + D_- \Phi^\nu \left( k^\mu E_{\mu\nu}
     - \omega_\nu \right) - D_- Y &=&0  &        (\ref{Veom})\mathrm{b} \\
y_- +\eta y_+ &=&0   &                                     (\ref{yans}) \\
v_- -\eta v_+ &=&0   &                                   (\ref{ysoln1}) \\
v_{--} + \eta v_{+-} - v_{++} - \eta v_{-+}  &=&0   &       (\ref{vpp})
\end{array} 
$$
The requirements for gauge invariance of each of
these conditions are as follows:
\vspace{-10pt}
\begin{enumerate}
\item Eqs.~(\ref{Qk0}), (\ref{Lg0}), (\ref{domega}), (\ref{LieR}),
      (\ref{LieQpi}), (\ref{LkpiE}), (\ref{piomega0}) and (\ref{Yeom}) are
      trivially gauge invariant.
\item (\ref{dty0}) is gauge invariant if $k^\mu \omega_\mu =0$, which
      follows from (\ref{piomega0}).
\item Eqs.~(\ref{Veom}) are gauge invariant due to (\ref{domega}).
\item Given (\ref{piomega0}), (\ref{yans}) is gauge invariant if
      $k^\mu \pi^\rho_{\,\,\,\mu} \pi^\nu_{\,\,\,\gamma}
     \pi^\lambda_{\,\,\,[\rho,\nu]}=0$, which follows from $\pi$-integrability,
     Eq.~(\ref{piinteg1}).
\item (\ref{ysoln1}) is gauge invariant due to $\epsilon_-
      - \eta \epsilon_+ =0$, Eq.~(\ref{epsrel}).
\item (\ref{vpp}) is gauge invariant due to the superpartner of
      $\epsilon_- - \eta \epsilon_+ =0$,
\beq
\epsilon_{--} +\eta\epsilon_{+-} -\epsilon_{++} - \eta\epsilon_{-+}=0 \, .  
\eeq{epp}
\end{enumerate}

\subsection{Summary and interpretation}

\begin{table}[ht]
\vspace{1cm}
\begin{tabular*}{\textwidth}{@{\extracolsep{\fill}}|ccc|}\hline
& $
\begin{array}{rcl@{\hspace{2cm}}l}
Q^\mu_{\,\,\,\nu} k^\nu&=&0 &                               (\ref{Qk0}) \\
{\cal L}_{k} g_{\mu\nu} &=&0 &                              (\ref{Lg0}) \\
{\cal L}_k B_{\mu\nu} - \omega_{[\nu,\mu]}&=&0  &        (\ref{domega}) \\ 
{\cal L}_k R^\mu_{\,\,\,\nu} &=&0  &                       (\ref{LieR}) \\
{\cal L}_k \pi^\mu_{\,\,\,\nu}
    = {\cal L}_k Q^\mu_{\,\,\,\nu} &=&0  &               (\ref{LieQpi}) \\
\pi^\mu_{\,\,\,\lambda} \left( \L_k B_{\mu\rho} \right)
         \pi^\rho_{\,\,\,\nu} &=& 0 &                     (\ref{LkpiE}) \\
\d_\tau y &=&0   &                                         (\ref{dty0}) \\
\pi^\mu_{\,\,\,\nu} \omega_\mu &=&0  &                 (\ref{piomega0}) \\
y_- +\eta y_+ &=&0   &                                     (\ref{yans}) \\
v_- -\eta v_+ &=&0   &                                   (\ref{ysoln1}) \\
\epsilon_- -\eta \epsilon_+ &=&0  &                      (\ref{epsrel}) \\
v_{--} + \eta v_{+-} - v_{++} - \eta v_{-+} &=& 0  &        (\ref{vpp}) \\
(\nabla K^\mu + \widehat S^\mu) - R^\mu_{\,\,\,\nu} (K)
  (\nabla K^\nu -\widehat S^\nu)  &=&0  &               (\ref{hatKSbc}) \\
Q^\mu_{\,\,\,\nu} \nabla K^\nu   &=&0  &                (\ref{QnablaK}) \\
\epsilon_{--} +\eta\epsilon_{+-}
  -\epsilon_{++} - \eta\epsilon_{-+} &=&0  &                (\ref{epp})  
\end{array} $ & \\\hline
\end{tabular*}
\caption{Boundary conditions for the ${\cal N}$=1
supersymmetric nonlinear sigma model to be consistent under T-duality.
We assume that the superconformality conditions listed in Table~\ref{N1bc}
are satisfied.}
\vspace{1cm}
\label{offshellbc}
\end{table}

Thus all boundary conditions are gauge invariant, and we summarise the
off-shell ones in  Table~\ref{offshellbc}.  The full set of boundary
conditions required for a consistent T-duality transformation of the
${\cal N}$=1 nonlinear sigma model are thus contained in
Tables~\ref{N1bc} and~\ref{offshellbc}.  The interpretation of the
superconformality conditions in Table~\ref{N1bc} is the same as in
\cite{ALZ2}, i.e., the D-branes in the original sigma model are
maximal integral submanifolds of the target space.  The conditions in
Table~\ref{offshellbc} are mostly the same as those required for the
bosonic model (see Section~\ref{bosinterp}) supplemented with boundary
conditions for the fermionic superpartner fields.  We therefore again
find that the D-brane in the original model
is an isometry invariant submanifold with
invariant metric and two-form, and that it loses one dimension as
we go to the dual model.

Eq.~(\ref{hatKSbc}) is just the ansatz we made for the relation
between left- and right-movers on the worldsheet, and (\ref{QnablaK})
is the usual Dirichlet condition.

\subsection{Boundary supersymmetry}
\label{susy}

Here we check that the parent action (\ref{gaugeact1}) is
supersymmetric on the boundary. To do this, we use the conditions in
Tables~\ref{N1bc} and~\ref{offshellbc} to verify that its
supersymmetry variation vanishes. The bulk part of (\ref{gaugeact1})
can be written in terms of a superfield Lagrangian $\L$,
$$
\widehat S_{\mathrm{bulk}} = \int d^2\xi \,\, D_+D_- \,\, \L \,.
$$
Applying the supersymmetry transformation (\ref{tfs}) we get
a pure boundary term,
$$
\delta_s \widehat S_{\mathrm{bulk}} = -i \int d\tau \,\, \epsilon^+
\,\, \left[ D_-\L + \eta  D_+\L \right] \,.
$$
The expansion of
$\delta_s \widehat S_{\mathrm{bulk}}$ in components reads
\beq
\delta_s \widehat S_{\mathrm{bulk}} = i \int d\tau \,\, \epsilon^+
\,\, \left[ \L_1 + \L_2 + \L_3 \right] \,,
\eeq{DDLvar1}
where
\ber
\nonumber \L_1 &\equiv&
\widehat\psi^\mu_+ \left( i\d_= X^\nu
     + k^\nu_{\,\,,\rho} \widehat\psi^\rho_- v_-
     + k^\nu v_{--} \right) E_{\mu\nu} \\
\nonumber && 
-\eta \left( i\d_\+ X^\mu
   + k^\mu_{\,\,,\rho} \widehat\psi^\rho_+ v_+ + k^\mu v_{++} \right)
   \widehat\psi^\nu_- E_{\mu\nu}\\
\nonumber && 
+\eta \widehat\psi^\mu_+ \left( F^\nu_{+-}
     + k^\nu_{\,\,,\rho} \widehat\psi^\rho_+ v_-
     + k^\nu v_{+-} \right) E_{\mu\nu}  \\
\nonumber && 
-\left( -F^\mu_{+-} + k^\mu_{\,\,,\rho} \widehat\psi^\rho_- v_+
                  + k^\mu v_{-+} \right) \widehat\psi^\nu_-  E_{\mu\nu}\\
&&
- \widehat\psi^\mu_+ \widehat\psi^\nu_-
     \left( \widehat\psi^\rho_- - k^\rho v_- \right)E_{\mu\nu,\rho}
- \eta \widehat\psi^\mu_+ \widehat\psi^\nu_-
     \left( \widehat\psi^\rho_+ - k^\rho v_+ \right)E_{\mu\nu,\rho}\,,
\label{L1def} \\
\nonumber && \\
\nonumber \L_2 &\equiv&
-  \left( \eta \widehat\psi^\rho_+  + \widehat\psi^\rho_- \right)
\omega_{\nu,\rho} 
    \left( \widehat\psi^\nu_+ v_- + \widehat\psi^\nu_- v_+ \right)  \\
\nonumber && 
+ \omega_{\nu} F^\nu_{+-}  \left( v_- - \eta v_+ \right)
-i\omega_\nu \d_= X^\nu v_+ -i\eta \omega_\nu \d_\+ X^\nu v_-   \\
&& 
+ \omega_\nu \left( \widehat\psi^\nu_+ - k^\nu v_+ \right)
             \left( v_{--} + \eta v_{+-} \right)
+ \omega_\nu \left( \widehat\psi^\nu_- - k^\nu v_- \right)
             \left(\eta v_{++} + v_{-+} \right) \,,
\label{L2def} \\
\nonumber && \\
\nonumber \L_3 & \equiv &
v_{--} y_+ + v_- y_{+-} + v_{-+} y_- - iv_+ \d_= y  \\
&& + \eta v_{++} y_-  - \eta v_+ y_{+-} + \eta v_{+-}y_+-i\eta v_- \d_\+ y \,.
\eer{L3def}
For the boundary $\widehat\psi\widehat\psi$-term
in (\ref{gaugeact1}) we use the
component form of the supersymmetry variation, Eq.~(\ref{compsusytr}).
This yields the following transformation,
\ber
\nonumber \delta_s \widehat S_{\widehat\psi\widehat\psi}
&=& i \int d\tau \,\, \epsilon^+ \,\, 
\left[ -i \d_\+ X^\mu \widehat\psi^\nu_+ B_{\mu\nu}
   -i \eta \d_= X^\mu \widehat\psi^\nu_- B_{\mu\nu}
+ \eta F^\mu_{+-} B_{\mu\nu}
 \left(\widehat\psi^\nu_+ - \eta \widehat\psi^\nu_- \right)
\right. \\
\nonumber && \left. 
+\left(\widehat\psi^\mu_+ + \eta \widehat\psi^\mu_-\right)
  B_{\mu\nu} k^\nu \left(v_{++} + \eta v_{-+}\right)
\right. \\
\nonumber && \left. 
-\half \left( \widehat\psi^\mu_+ \widehat\psi^\nu_+ 
             +\widehat\psi^\mu_- \widehat\psi^\nu_-\right)
B_{\mu\nu,\rho} \left(\widehat\psi^\rho_+ + \eta \widehat\psi^\rho_-\right)
\right. \\
\nonumber && \left.
+ v_+ \left( \widehat\psi^\mu_+ \widehat\psi^\nu_+ 
             +\widehat\psi^\mu_- \widehat\psi^\nu_-\right)
      \left( \L_k B_{\mu\nu} - B_{\rho\nu} k^\rho_{\,\,,\mu} \right)
\right. \\
&& \left. 
+\eta v_+ \widehat\psi^\mu_+ \widehat\psi^\nu_-
      \left( \L_k B_{\mu\nu} - B_{\mu\nu,\rho} k^\rho \right)
\right] .
\eer{psipsivar}
The parent action is supersymmetric if $\delta_s \widehat
S_{\mathrm{bulk}} + \delta_s \widehat S_{\widehat\psi\widehat\psi}=0$.
We start by noting that $\L_3$ vanishes when we use (\ref{dty0}), (\ref{yans})
and (\ref{ysoln1}).  Similarly, $\L_2$ vanishes when we
use (\ref{LkpiE}), (\ref{piomega0}) and (\ref{ysoln1}).  We are thus
left with $\L_1$
and $\delta_s \widehat S_{\widehat\psi\widehat\psi}$.  We first collect the
$\d_= X^\mu$-terms and substitute the fermionic ansatz (\ref{fermans3})
and its bosonic superpartner (\ref{dmXans}).
The $\d_= X^\mu$-terms in (\ref{L1def}) plus (\ref{psipsivar}) read
\beq
i \d_= X^\mu \left(
E_{\nu\mu} \widehat\psi^\nu_+ - \eta B_{\mu\nu} \widehat\psi^\nu_-
\right) ,
\eeq{dmXterms}
and after using the ansatz the total supersymmetry variation contains
five different types of terms: $\d_\+ X^\mu$-terms,
$F^\mu_{+-}$-terms, $(v_{++} + \eta v_{-+})$-terms, $v_+
\widehat\psi^\mu_+ \widehat\psi^\nu_+$-terms, and $\widehat\psi^\mu_+
\widehat\psi^\nu_+ \widehat\psi^\rho_+$-terms.
It is straightforward to verify that each type of term vanishes
separately when we use conditions (\ref{RgRcond1}),
(\ref{pigQ}), (\ref{piinteg1}), (\ref{piEpiR})
and (\ref{LieR}). A crucial relation in this computation follows
from the four conditions (\ref{RgRcond1}), (\ref{pigQ}),
(\ref{piinteg1}) and (\ref{piEpiR}), and reads
\beq
R^\mu_{\,\,\,\rho} B_{\mu\lambda} R^\lambda_{\,\,\,\nu} 
= -B_{\rho\nu} - E_{\rho\lambda} R^\lambda_{\,\,\,\nu}
+E_{\nu\lambda} R^\lambda_{\,\,\,\rho}  \, .
\eeq{RBR}
We have thus shown that the total variation $\delta_s \widehat
S_{\mathrm{bulk}} + \delta_s \widehat S_{\widehat\psi\widehat\psi}$
vanishes when we use the boundary conditions listed
in Tables~\ref{N1bc} and~\ref{offshellbc}.  Hence the parent action is
indeed supersymmetric given all the conditions derived for consistency
and gauge invariance.

We also need to check that the superpartner of each of the boundary
conditions holds, since the supersymmetry algebra must close. The ansatz
(\ref{hatKSbc}) and the Dirichlet condition
(\ref{QnablaK}) are both manifestly supersymmetric,
and for each of the other boundary conditions whose superpartner is
nontrivial, its superpartner is included in the set of required
conditions.  This is obvious when we list them in
superpartner pairs:
$$
\begin{array}{rcl@{\hspace{2cm}}l}
\d_\tau y &=&0  &                               (\ref{dty0}) \\
y_- +\eta y_+ &=&0                          &   (\ref{yans}) \\
&&& \\
v_- -\eta v_+ &=&0  &                         (\ref{ysoln1}) \\
v_{--} + \eta v_{+-} - v_{++} - \eta v_{-+}
       &=& 0  &                                  (\ref{vpp}) \\
&&& \\
\epsilon_- -\eta \epsilon_+ &=&0   &          (\ref{epsrel}) \\
\epsilon_{--} + \eta \epsilon_{+-} - \epsilon_{++} -
                        \eta \epsilon_{-+} &=& 0  &(\ref{epp})
\end{array}
$$
The rest of the conditions in Tables~\ref{N1bc} and~\ref{offshellbc}
are trivially supersymmetric, hence we have shown that the gauged
model is completely supersymmetric on the boundary.

\Section{Boundary couplings}
\label{couplings}

Interactions are in general described by a path integral
of the form
$$
 \int D\Phi \,\, ({\cal V} {\cal V} ... {\cal V}) \,\, e^{-S} ,
$$
where the ${\cal V}$'s denote vertex operators and $S$ is the worldsheet
action.  The open string may couple to fields on the brane, giving rise
to interactions that modify the action $S$
by boundary terms. One possible such perturbation
is the standard $U(1)$ boundary coupling
(using 1D superfields) \cite{BorlafLozano,Tseytlin,Andreev1,Andreev2}
\beq
S_{U(1)} = \int d\tau d\theta \,\, A_{\mu}(K)DK^\mu = \int d\tau
 \left ( i A_\mu(X) \d_\tau X^\mu -\frac{1}{2} F_{\mu\nu} (X)
 \Psi^\mu \Psi^\nu \right ) \, ,
\eeq{Acoupling}
where $F_{\mu\nu} \equiv \d_{[\mu} A_{\nu]}$.
The fact that the $A$-field lives on the brane is
expressed by the condition $A_\mu Q^\mu_{\,\,\,\nu}=0$.
Another boundary perturbation corresponds to
transverse fluctuations of the brane \cite{Sfluct},
\beq
S_{fluct} =
\int d\tau d\theta \,\, Y_{\mu}(K) S^\mu = \int d\tau \left (Y_\mu(X)
  (\eta F_{+-}^\mu - i\partial_\sigma X^\mu) + Y_{\mu,\nu} \Psi^\nu
  \widehat{\Psi}^\mu \right ) \, ,
\eeq{Ycoupl}
where $Y_\mu$ is an auxiliary scalar field.
Moreover, introducing the auxiliary spinor superfield
$\Gamma= d + \theta Z$, one can write down a tachyonic vertex
with tachyon potential ${\cal T}$ \cite{Stach},
\beq
S_{tach} =
 \int d\tau d\theta \,\,({\cal T}(K)\Gamma + \Gamma D \Gamma)
 =  \int d\tau \left (
 \frac{\d {\cal T}}{\d X^\mu} \Psi^\mu d
 + {\cal T}(X)Z +Z^2 - i d \d_\tau d \right ) \, .
\eeq{tachyon}
There are also boundary perturbations corresponding
to massive open string states \cite{Smass},
\beq
S^{(1)}_{mass} =
 \int d\tau d\theta \,\, A_{\mu\nu}(K) D^2K^\mu DK^\nu \, ,
\eeq{masscoupling2}
where $A_{\mu\nu}$ is a massive symmetric tensor field, or
\beq
S^{(2)}_{mass} =
 \int d\tau d\theta \,\, A_{\mu\nu\rho}(K) DK^\mu DK^\nu DK^\rho \, ,
\eeq{masscoupling}
for a massive general tensor $A_{\mu\nu\rho}$.

Note that the boundary interactions $S_{U(1)}$, $S_{fluct}$,
$S_{tach}$, $S^{(1)}_{mass}$ and $S^{(2)}_{mass}$ are individually
supersymmetric, whereas the nonlinear sigma model (\ref{goodaction}) is
supersymmetric only subject to the boundary conditions
(\ref{fermans})--(\ref{ALZ320}). Thus the boundary interactions are
suitable for supersymmetric perturbation theory in the sense that the
action (\ref{goodaction}) produces the supersymmetric propagators and
the perturbations (\ref{Acoupling})--(\ref{masscoupling}) give the
supersymmetric vertices.

When modifying the worldsheet action with boundary interactions, we
must check that the new action is still isometry invariant as well as
consistent with T-duality. Since we know that these criteria are
fulfilled by the action (\ref{goodaction}) separately, and we know how to
gauge it in a consistent way, it remains to
check that the boundary perturbations $S_{U(1)}$, $S_{fluct}$,
$S_{tach}$, $S^{(1)}_{mass}$ and $S^{(2)}_{mass}$ are independently isometry
invariant and that they can be gauged.

\subsection{Isometry invariance}

Under the isometry transformation (\ref{1Dtrans})
the $U(1)$ coupling (\ref{Acoupling}) transforms as
(recall that $D\epsilon^A=0$)
$$
\delta_{k} S_{U(1)} =
\epsilon^A \int d\tau d\theta \,\, (\L_{k_A} A_{\mu}) \, DK^\mu \, .
$$
For this to vanish, the Lie derivative must satisfy
\beq
\pi^\mu_{\,\,\,\nu} \L_{k_A} A_{\mu} = \d_\nu C_A \, ,
\eeq{LAU1}
for some scalar field $C_A$.
The transformation acts identically on $S_{fluct}$, resulting in
the condition
\beq
\pi^\mu_{\,\,\,\nu} \L_{k_A} Y_{\mu} = \d_\nu C_A \, .
\eeq{LY}
For the tachyonic vertex we assume that $\Gamma$ is annihilated by
an isometry transformation, $\delta_{k_A} \Gamma=0$, hence $S_{tach}$
transforms as
$$
\delta_{k} S_{tach} =
\epsilon^A \int d\tau d\theta \,\, (\L_{k_A} {\cal T})\, \Gamma \, ,
$$
requiring
\beq
\L_{k_A} {\cal T} =0
\eeq{LT}
in order to vanish.
The massive case is analogous to that of $S_{U(1)}$ and we have
$$
\delta_{k} S^{(1)}_{mass} =
i \epsilon^A \int d\tau d\theta \,\,
(\L_{k_A} A_{\mu\nu}) \, \d_\tau K^\mu \, DK^\nu \, ,
$$
implying
\beq
\pi^\mu_{\,\,\,\rho} (\L_{k_A} A_{\mu\nu}) \pi^\nu_{\,\,\,\lambda}
 = \d_\rho \d_\lambda C_A \, .
\eeq{LAmass1}
For $S^{(2)}_{mass}$ we have
$$
\delta_{k} S^{(2)}_{mass} =
\epsilon^A \int d\tau d\theta \,\,
(\L_{k_A} A_{\mu\nu\rho}) \, DK^\mu \,DK^\nu \,DK^\rho \, ,
$$
implying
$$
\pi^\mu_{\,\,\,\sigma}  \pi^\nu_{\,\,\,\lambda} \pi^\rho_{\,\,\,\gamma}
\L_{k_A} A_{\mu\nu\rho}  = \d_\sigma \d_\lambda \d_\gamma C_A \, .
$$
Note that the last statement effectively declares that
\beq
\pi^\mu_{\,\,\,\sigma}  \pi^\nu_{\,\,\,\lambda} \pi^\rho_{\,\,\,\gamma}
\L_{k_A} A_{\mu\nu\rho}  = 0 \, ,
\eeq{LAmass2}
since only the antisymmetric piece of $A_{\mu\nu\rho}$ is relevant
in $S^{(2)}_{mass}$.

In conclusion, the boundary couplings must satisfy
(\ref{LAU1})--(\ref{LAmass2}) for the resulting worldsheet action to
be isometry invariant.

\subsection{T-duality of vertices}

We now promote the boundary actions $S_{U(1)}$, $S_{fluct}$,
$S_{tach}$, $S^{(1)}_{mass}$ and $S^{(2)}_{mass}$ to parent actions by
adding fields to make them invariant under local gauge
transformations (after gauging a $U(1)$ isometry).
Starting with the $U(1)$ coupling and taking
$D\epsilon \neq 0$, we find that the action
$$
\widehat S_{U(1)} = \int d\tau d\theta \,\,
 A_{\mu}(K) \,\nabla K^\mu 
$$
is gauge invariant if
$$
\nabla K^\mu \equiv D K^\mu + k^\mu V_1 \, ,
\quad\quad \delta_k V_1 = -D\epsilon \, .
$$
This is in complete analogy with
the gauged form of the nonlinear sigma model.
The parent action for $S_{fluct}$ is almost identical:
$$
\widehat S_{fluct} = \int d\tau d\theta \,\,
 Y_{\mu}(K) \,\widehat S^\mu \, ,
$$
where
$$
\widehat S^\mu \equiv S^\mu + k^\mu V_2 \, ,
\quad\quad \delta_k V_2 = -D\epsilon \, .
$$
The tachyonic action, however, does not receive any corrections since
a gauge transformation of $S_{tach}$ leaves no $D\epsilon$-terms.
Finally, the massive vertices are analogous to the $U(1)$ coupling,
producing the following parent actions:
$$
\widehat S^{(1)}_{mass} = \int d\tau d\theta \,\,
 A_{\mu\nu}\,  \nabla_\tau K^\mu \, \nabla K^\nu \, ,
$$
where
$$
\nabla_\tau K^\mu \equiv i\d_\tau K^\mu + k^\mu DV_1 \, ,
$$
and
$$
\widehat S^{(2)}_{mass} = \int d\tau d\theta \,\,
 A_{\mu\nu\rho} \,\nabla K^\mu \,\nabla K^\nu \,\nabla K^\rho \, .
$$

One may now write the boundary interactions as
$$
\int D\Phi DV_{\pm} \,\, ({\cal V} {\cal V} ... {\cal V} )
 \,\, e^{-S[\Phi,V]} \,,
$$
where $S[\Phi,V]$ is the gauged action (\ref{gaugeact1}) plus any of
the boundary interactions $\widehat S_{U(1)}$, $\widehat S_{fluct}$,
$\widehat S_{tach}$, $\widehat S^{(1)}_{mass}$ or $\widehat
S^{(2)}_{mass}$ .
The T-dual vertices\footnote{For another discussion of
T-duality of boundary couplings, see \cite{Hassan:2003uq}.}
are obtained by plugging in the
bulk values for the fields $V_{1,2}$ and fixing the gauge.

\Section{T-duality of boundary conditions}
\label{Tdualbc}

Up to this point we have obtained the action T-dual to the
supersymmetric nonlinear sigma model, and the boundary conditions
necessary to ensure a consistent dualisation procedure.
There is one more aspect to investigate: How do the boundary
conditions themselves transform under T-duality? Are there any further
conditions arising from such a consideration?  This is a topic for
extensive analysis and we will only comment on it briefly here,
deferring a more detailed discussion to subsequent work.

We start by examining the transformation properties of the worldsheet
fields.  We look at the bosonic model here; the supersymmetric case is
completely analogous and yields the same results.  The isometry group is
again $U(1)$ generated by a Killing vector $k^\mu$.
The worldsheet fields transform under T-duality according to the
equations of motion (\ref{Aeom}) after fixing the gauge. The resulting 
transformation can be written in a compact form in
terms of matrices $\Qpm$ acting on the worldsheet fields
(schematically) as follows \cite{Hassan},
\beq
\left( \begin{array}{c}
 \d_\+\widetilde X^0 \\
 \d_\+\widetilde X^n
\end{array} \right)
= \Qp \left( \begin{array}{c}
 \d_\+ X^0 \\
 \d_\+ X^n
\end{array} \right) ,\,\,\,\,\,\,\,\,\,\,\,\,\,\,\,\,\,\,
\left( \begin{array}{c}
 \d_=\widetilde X^0 \\
 \d_=\widetilde X^n
\end{array} \right)
= \Qm \left( \begin{array}{c}
 \d_= X^0 \\
 \d_= X^n
\end{array} \right) ,
\eeq{QpmdX}
where the matrices $\Qpm$ are defined as
\beq
 \Qp \equiv \left ( \begin{array}{cc}
 - E_{00} & - E_{n0} -\omega_n \\
  0 & \bid 
 \end{array} \right ),\,\,\,\,\,\,\,\,\,\,\,\,\,\,\,\,\,\,
 \Qm \equiv \left (\begin{array}{cc}
  E_{00}  & E_{0n} -\omega_n \\
  0  & \bid
\end{array} \right ) ,
\eeq{QQmatr}
and $\bid$ is the identity matrix.

We know that conformal invariance requires the boundary condition
$$
\begin{array}{l@{\hspace{3cm}}l}
\d_= X^\mu = R^\mu_{\,\,\,\nu} \d_\+ X^\nu\,, & (\ref{bosans}) \\
\end{array}
$$
with $R^\mu_{\,\,\,\nu}$ satisfying
$$
\begin{array}{l@{\hspace{3.2cm}}l}
g_{\mu\nu} = R^\rho_{\,\,\,\mu} g_{\rho\sigma} R^\sigma_{\,\,\,\nu}\,.
& (\ref{RgRcond1})
\end{array}
$$
One can perform a direct T-duality transformation of these conditions
by applying (\ref{QpmdX}), and because the dual action should also be
conformally invariant the dual counterparts of (\ref{bosans}) and
(\ref{RgRcond1}) must be identical in form, i.e., we expect
$$
\d_= \widetilde X^\mu = \widetilde R^\mu_{\,\,\,\nu} \d_\+ \widetilde X^\nu\,, 
\quad\quad
\widetilde g_{\mu\nu} = \widetilde R^\rho_{\,\,\,\mu}
  \widetilde g_{\rho\sigma} \widetilde R^\sigma_{\,\,\,\nu}\,.
$$
The transformation of (\ref{bosans}) reads (schematically)
$$
\d_= \widetilde X = \Qm \, R \, \Qp^{-1} \, \d_\+ \widetilde X\,,
$$
leading us to the conclusion that
$$
\widetilde R = \Qm \, R \, \Qp^{-1} \,.
$$
Given this relation, the transformation of (\ref{RgRcond1}),
$$
(\Qp^{-1})^t \, g \, \Qp^{-1}
 = \widetilde R^t \, (\Qm^{-1})^t \, g \, \Qm^{-1} \, \widetilde R\,,
$$
implies
$$
\widetilde g = (\Qp^{-1})^t \, g \, \Qp^{-1} = (\Qm^{-1})^t \, g\,\Qm^{-1}\,,
$$
which is just a different way of writing Buscher's rules (\ref{Buscher}).

For a complete treatment, the rest of the boundary conditions in
Table~\ref{N1bc} should be similarly analysed, and the T-duality
transformation for the projectors $\pi^\mu_{\,\,\,\nu}$ and
$Q^\mu_{\,\,\,\nu}$ should be derived. This will be done elsewhere.

\Section{Conclusions}
\label{conclusions}

We have shown how to gauge a $U(1)$ isometry in the ${\cal N}$=1
supersymmetric nonlinear sigma model with boundaries in a consistent
way to obtain its T-dual via a gauge invariant parent action. We have
derived the most general boundary conditions that must be satisfied by
the worldsheet fields in order for the duality transformation to be
consistent.  The bosonic model was examined first, and we found that
it requires a restricted form of integrability, which can be
interpreted in terms of a D-brane whose projection onto the isometry
direction in the original model is an integral submanifold of the
target space.  The D-brane is also an isometry invariant submanifold
equipped with invariant geometrical data.

For the supersymmetric sigma model we derived the conditions necessary
for the isometry group to be a symmetry of the original action, and then
defined the appropriate parent action, finding the boundary conditions
required for it to be consistent at the level of field equations as
well as gauge invariant and supersymmetric. The resulting conditions
are the same as for the ungauged sigma model (see Table~\ref{N1bc}),
plus a number of restrictions on the background and the added gauge
fields and lagrange multipliers (see Table~\ref{offshellbc}). The
interpretation of the conditions in Table~\ref{N1bc} is exactly that
given in \cite{ALZ2}, i.e., the D-brane in the original model is a
maximal integral submanifold
of the target space. The conditions in Table~\ref{offshellbc} encode the
transition of the Neumann isometry direction in the original model to
a Dirichlet direction in the dual model, together with the requirement
that the background should be independent of the isometry direction
for the procedure to work.  They imply that the D-brane is again an
invariant submanifold with invariant metric and two-form.

We moreover showed how to incorporate couplings to boundary states in
the T-duality transformation, and briefly discussed how the boundary
conditions themselves transform under T-duality, an issue that deserves
a more detailed analysis.

For a complete picture, future studies should include
sigma models with more general abelian and
nonabelian isometry groups, as well as models
without isometries. In the latter case the gauging procedure is
unsuitable for performing T-duality, and one has to resort to other,
more involved methods, e.g., Poisson-Lie T-duality.  In addition to
classical T-duality, quantum aspects should also be investigated. This
involves extending the analysis to string worldsheets of more
complicated topology.

\bigskip

\bigskip

{\bf Acknowledgements}:
We are grateful to Martin Ro\v{c}ek for discussions and comments.
CA acknowledges support by Deutsche Forschungsgemeinschaft (DFG).
UL acknowledges support by VR grant 650-1998368.
MZ acknowledges support by EU grant MEIF\_CT-2004-500267.

\appendix

\Section{(1,1) supersymmetry}
\label{a:11susy}

Throughout the paper we use $\mu,\nu,...$ as spacetime indices,
$(\+,=)$ as worldsheet indices (in lightcone coordinates $\xi^\pm
\equiv \tau \pm \sigma$, where $\tau$, $\sigma$ are the usual
worldsheet coordinates), and $(+,-)$ as two-dimensional spinor
indices.  We also use superspace conventions where the pair of spinor
coordinates are labelled $\th^{\pm}$, and the covariant derivatives
$D_\pm$ and supersymmetry generators $Q_\pm$ satisfy
\ber
D^2_+ &=&i\d_\+, \quad
D^2_- =i\d_= , \quad \{D_+,D_-\}=0 , \cr
Q_\pm &=& -D_\pm+2i\th^{\pm}\d_{\pp} ,
\eer{alg}
where $\d_{\pp}=\partial_\tau \pm \partial_\sigma$.
In terms of the covariant derivatives, a supersymmetry transformation of
a superfield $\P$ is given by
\ber
\delta \P &\equiv & (\epsilon^+Q_++\epsilon^-Q_-)\P \cr
&=& -(\epsilon^+D_++\epsilon^-D_-)\P
+2i(\epsilon^+\th^+\d_\++\epsilon^-\th^-\d_=)\P ,
\eer{tfs}
where $\epsilon^\pm$ are the left- and right-moving supersymmetry parameters.
The components of a superfield $\P$ are defined via projections as
follows,
\ber
\P|\equiv X, \quad D_\pm\P| \equiv \p_\pm, \quad D_+D_-\P|\equiv F_{+-}
,
\eer{comp}
where a vertical bar denotes ``the $\th =0$ part of ''.
Thus, in
components, the $(1,1)$ supersymmetry transformations are given by
\beq
\left \{ \begin{array}{l}
 \delta X^\mu = - \epsilon^{+} \psi_+^\mu - \epsilon^- \psi_-^\mu  \\
 \delta \psi_+^\mu =  -i\epsilon^+ \d_{\+}X^\mu - \epsilon^- F^\mu_{-+}  \\
\delta \psi_-^\mu  = -i \epsilon^- \d_{=} X^\mu - \epsilon^+ F_{+-}^\mu  \\
\delta F^\mu_{+-} = - i \epsilon^+ \d_{\+} \psi_-^\mu +
 i \epsilon^- \d_= \psi_+^\mu 
\end{array} \right .
\eeq{compsusytr}

\Section{1D superfield formalism}
\label{a:1D}

One may view the 2D supersymmetry algebra as a combination of two 1D
algebras (for similar considerations see \cite{Hori,Sevrin}). To see this, we
rewrite the $(1,1)$ supersymmetry algebra in terms of 1D
supermultiplets.  Defining $\epsilon\equiv \epsilon^-$ and
assuming that $\epsilon^+ = \eta \epsilon$, (\ref{compsusytr}) becomes
\beq 
\left \{
\begin{array}{l}
  \delta X^\mu = - \epsilon (\eta \psi_+^\mu + \psi_-^\mu) \\
  \delta (\eta \psi_+^\mu + \psi_-^\mu) = 
 -2i\epsilon \partial_\tau X^\mu \\
 \delta ( \psi_-^\mu - \eta \psi_+^\mu ) = -2\epsilon
 ( \eta F_{+-}^\mu - i\partial_\sigma X^\mu ) \\
 \delta (\eta F_{+-}^\mu -i\partial_\sigma X^\mu ) =
 -i \epsilon \partial_\tau (  \psi_-^\mu - \eta \psi_+^\mu)
\end{array} \right.  
\eeq{susyrew1}
Introducing a new notation for the following
combinations of fields,
\beq
\Psi^\mu \equiv \frac{1}{\sqrt{2}}(\psi_-^\mu + \eta \psi_+^\mu )\,,
\,\,\,\,\,\,\,\,\,
\widehat{\Psi}^\mu \equiv \frac{1}{\sqrt{2}}
( \psi_-^\mu -\eta \psi_+^\mu )\,,\,\,\,\,\,\,\,\,\,
f^\mu \equiv \eta F_{+-}^\mu -i \partial_\sigma X^\mu\,,
\eeq{newnot}
and redefining $\epsilon= \sqrt{2}\epsilon$,
the algebra (\ref{susyrew1}) takes on the simple form
\beq 
\left \{
\begin{array}{l}
  \delta X^\mu = - \epsilon \Psi^\mu \\
  \delta \Psi^\mu =  - i\epsilon \partial_\tau X^\mu \\
  \delta \widehat{\Psi}^\mu = -\epsilon  f^\mu \\
  \delta f^\mu =  - i \epsilon \partial_\tau \widehat{\Psi}^\mu 
\end{array} \right.  
\eeq{susyrew2}
Clearly, (\ref{susyrew2}) is a decomposition of the
2D algebra into two 1D supermultiplets.
We introduce a 1D superfield notation for these multiplets,
\beq
 K^\mu = X^\mu +\theta \Psi^\mu,\,\,\,\,\,\,\,\,\,\,\,\,\,\,\,\,
 S^\mu = \widehat{\Psi}^\mu + \theta f^\mu,
\eeq{1dsusyf}
where $\theta$ is the single Grassmann coordinate of the respective 1D
superspace, and the corresponding 1D superderivative is now $D$,
satisfying $D^2=i\partial_\tau$.

\Section{The Lie derivative}
\label{a:geom}

The Lie derivative with respect to a vector $k$, acting on an
$(l,s)$-tensor $T$, is defined, in coordinate components, as
\beq
({\cal L}_k T)^{\mu_1\mu_2 ...\mu_l}_{\,\,\,\,\,\,\nu_1\nu_2...\nu_s} =
  T^{\mu_1\mu_2 ...\mu_l}_{\,\,\,\,\,\,\nu_1\nu_2...\nu_s,\rho} k^\rho -
 T^{\rho\mu_2 ...\mu_l}_{\,\,\,\,\,\,\nu_1\nu_2...\nu_s}
 k^{\mu_1}_{,\rho} - ...
 + T^{\mu_1\mu_2 ...\mu_l}_{\,\,\,\,\,\,\rho\nu_2...\nu_s}
 k^\rho_{,\nu_1} + ...
\eeq{deflie}

\Section{Supercurrents of the gauged model}
\label{gaugecur}

Here we write down the currents corresponding to superconformal
invariance of the parent action (\ref{gaugeact1}) for
the special case where $\L_k B_{\mu\nu}=0$. They are required to satisfy
the following boundary conditions,
\beq
T_{++}-T_{--} =0 \,, \,\,\,\,\,\,\,\,\,\,\, G_+ - \eta G_- = 0\,,
\eeq{TG}
where the components are defined as follows,
\beq
G_{+} = T_{\+}^-| \, ,\,\,\,\,\,\,\,\,\,\,\,
G_{-} = T_{=}^+|\, ,
\eeq{Gcomp}
\beq
T_{++} = -iD_{+} T_{\+}^-| \, , \,\,\,\,\,\,\,\,\,\,\,
T_{--} = - iD_{-} T_{=}^+| \,,
\eeq{Tcomp}
with supercurrent components given by
\beq
\begin{array}{rcr}
T_{\+}^- & \equiv &
- \frac{i}{2}D_+({\cal D}_+\Phi^\mu{\cal
    D}_+\Phi^\nu E_{\mu\nu}) +{\cal D}_+\Phi^\mu{\cal D}_{\+}\Phi^\nu
  E_{\mu\nu} \,\, , \\
T_{=}^+ & \equiv & \frac{i}{2} D_-({\cal D}_-\Phi^\mu{\cal
      D}_-\Phi^\nu E_{\mu\nu}) + {\cal D}_=\Phi^\mu{\cal D}_-\Phi^\nu
    E_{\mu\nu} \,\, ,
\end{array}
\eeq{supercur1}
or, expanding the first term in each current,
\beq
\begin{array}{rcr}
T_{\+}^- &=& \D_{+}\Phi^\mu \D_{\+} \Phi^\nu g_{\mu\nu} - \frac{i}{2}
 \D_{+}\Phi^\mu \D_{+} \Phi^\nu D_{+} \Phi^\rho E_{\mu\nu , \rho} \\
T_{=}^+ &=& \D_{-}\Phi^\mu \D_{=} \Phi^\nu g_{\mu\nu} + \frac{i}{2}
 \D_{-}\Phi^\mu \D_{-} \Phi^\nu D_{-} \Phi^\rho E_{\mu\nu , \rho}\, .
\end{array}
\eeq{supercur2}
The supercurrents are conserved,
\beq
 D_{+}T_{=}^+ =0\,,\,\,\,\,\,\,\,\,\,\,\,\,\,\,\, D_{-}T_{\+}^- =0 \,.
\eeq{conservlaws}
Here we have used the following notation,
\ber
{\cal D}_\pm \Phi^\mu &\equiv & D_\pm \Phi^\mu+\delta^{\mu 0} V_\pm~, \\
{\cal D}_{\pp} \Phi^\mu &\equiv & \d_{\pp} \Phi^\mu -i\delta^{\mu 0} V_{\pp}~,
\eer{calD}
where
\beq
V_\+ \equiv D_+ V_+\,, \quad\quad V_= \equiv D_-V_-~,
\eeq{Vdef}
and the derivatives satisfy the following relations,
$$
\begin{array}{r@{\hspace{1cm}}r}
D_{\pm}^2 = i \d_{\pp}\, , & \D_{\pm}^2 = i \D_{\pp} \,\, , \\
\{ D_+ , D_- \} =0\,,  & \{ \D_+ , \D_- \} =0 \,\, , \\
D_+ \D_+ \Phi^\mu = i \D_\+ \Phi^\mu \,,  & 
D_- \D_- \Phi^\mu = i \D_= \Phi^\mu \,\, , 
\end{array}
$$
\beq
\D_\+ \D_\pm \Phi^\mu = \D_\pm \D_\+ \Phi^\mu
= \d_\+ \D_\pm \Phi^\mu \,\, .
\eeq{Drelns2}

The currents (\ref{supercur2}) can be written on the more symmetric form
($H_{\mu\nu\rho}$ is the field strength of the B-field)
\beq
\begin{array}{rcr}
T_{\+}^- &=& \D_{+}\Phi^\mu \D_{\+} \Phi^\nu g_{\mu\nu} - \frac{i}{3}
 \D_{+}\Phi^\mu \D_{+} \Phi^\nu \D_{+} \Phi^\rho H_{\mu\nu\rho} \,\, ,\\
T_{=}^+ &=& \D_{-}\Phi^\mu \D_{=} \Phi^\nu g_{\mu\nu} + \frac{i}{3}
 \D_{-}\Phi^\mu \D_{-} \Phi^\nu \D_{-} \Phi^\rho H_{\mu\nu\rho} \,\, ,
\end{array}
\eeq{supercur3}
so we see that they are obtained from the ``ungauged'' supercurrents
(see \cite{ALZ2})
simply by replacing ordinary superderivatives $D_\pm$ with covariant
ones, $\D_\pm$. These gauged currents reduce to those of the original and
dual sigma models respectively, when we integrate out the
appropriate fields ($Y$ and $V_\pm$, respectively).

\end{document}